\documentclass[twocolums, longauth]{aa}
\usepackage{graphicx}

\usepackage{enumitem}

\def\deg{^{\circ}}
\usepackage{siunitx}
\usepackage{txfonts}

\renewcommand{\arraystretch}{1.5}

\begin{document}

\title{The $\beta$ Pictoris system: Setting constraints on the planet and the disk structures at mid-IR wavelengths with NEAR\thanks{Based on data collected at the European Southern Observatory under programs 60.A-9107(K), 095.C-0425(A), and 60.A-9234(A).}}

\institute{
LESIA, Observatoire de Paris, Univ. Paris Cit\'{e}, Univ.~PSL, CNRS, Sorbonne Univ., 5 pl. Jules Janssen, 92195 Meudon, France\label{inst:LESIA} %
\and 
National Astronomical Observatory of Japan, Subaru Telescope, 650 North A'oh\=ok\=u Place, Hilo, HI 96720, U.S.A.\label{inst:Subaru} %
\and 
Department of Physics and Astronomy, University College London, London, United Kingdom\label{inst:UCL} 
\and 
Universit\'e Paris-Saclay, Universit\'{e} Paris Cit\'{e} CEA, CNRS, AIM, 91191, Gif-sur-Yvette, France \label{inst:CEA} %
\and 
Instituto de Astrof\'isica de Andaluc\'ia, CSIC, Glorieta de la Astronom\'ia s/n, 18008, Granada, Spain \label{inst:spain} %
\and 
Universit\'{e} Grenoble Alpes, CNRS, IPAG, 38000 Grenoble, France \label{inst:IPAG} %
\and
IRAP, Universit\`{e} de Toulouse, CNRS, UPS, Toulouse, France \label{inst:IRAP}%
\and 
Leiden Observatory, Leiden University, P.O. Box 9513, 2300 RA Leiden, The Netherlands \label{inst:Leiden} %
\and 
STAR Institute, Universit\'e de Li\`ege, All\'ee du Six Ao\^ut 19c, 4000 Li\`ege, Belgium \label{inst:Belgium} %
\and
Universit\'{e} de Lyon, Universit\'{e} Lyon1, ENS de Lyon, CNRS, Centre de Recherche Astrophysique de Lyon UMR 5574, 69230 Saint-
Genis-Laval, France \label{inst:CRAL}
\and
Max Planck Institut für Astronomie, Königstuhl 17, 69117 Heidelberg, Germany \label{inst:MPIA}%
\and
Laboratoire J.-L. Lagrange, Universit \'e Cote d’Azur, CNRS, Observatoire de la Cote d’Azur, 06304 Nice, France \label{inst:OCA}
\and 
Centro de Astrobiolog\'{\i}a (CAB), CSIC-INTA, ESAC Campus, Camino bajo del Castillo s/n, E-28692 Villanueva de la Ca\~nada, Madrid, Spain \label{inst:spain2}
\and
Astrobiology Center of NINS, 2-21-1 Osawa, Mitaka, Tokyo 181-8588, Japan\label{inst:ABC}
\and
Steward Observatory, University of Arizona, Tucson, AZ 85721, USA\label{inst:UoA}
\and
College of Optical Sciences, University of Arizona, Tucson, AZ 85721, USA\label{inst:UoAO}
\and
Aix Marseille Univ, CNRS, CNES, LAM, Marseille, France \label{inst:LAM} %
\and 
Lund Observatory, Department of Astronomy and Theoretical Physics, Lund University, Box 43, 221 00 Lund, Sweden\label{inst:sweden}
\and
European Southern Observatory, Karl-Schwarzschild-Str. 2, 85748, Garching, Germany\label{inst:ESO}
}

\author{
    Nour Skaf\inst{\ref{inst:LESIA},\ref{inst:Subaru},\ref{inst:UCL}} \and%
	Anthony Boccaletti\inst{\ref{inst:LESIA}} \and%
	Eric Pantin\inst{\ref{inst:CEA}} \and%
	Philippe Thebault\inst{\ref{inst:LESIA}} \and%
	Quentin Kral\inst{\ref{inst:LESIA}} \and%
	Camilla Danielski \inst{\ref{inst:spain}}\and
	Raphael Galicher \inst{\ref{inst:LESIA}} \and
	Julien Milli \inst{\ref{inst:IPAG}}\and
	Anne-Marie Lagrange \inst{\ref{inst:LESIA}}\and
	Cl\'{e}ment Baruteau \inst{\ref{inst:IRAP}} \and
	Matthew Kenworthy \inst{\ref{inst:Leiden}} \and 
	Olivier Absil\inst{\ref{inst:Belgium}} \and 
	Maud Langlois \inst{\ref{inst:CRAL}}\and
    Johan Olofsson \inst{\ref{inst:MPIA}} \and 
	Gael Chauvin \inst{\ref{inst:OCA}}\and 
	Nuria Huelamo \inst{\ref{inst:spain2}} \and
	Philippe Delorme \inst{\ref{inst:IPAG}} \and 
	Benjamin Charnay\inst{\ref{inst:LESIA}} \and%
	Olivier Guyon  \inst{\ref{inst:Subaru},\ref{inst:ABC},\ref{inst:UoA}, \ref{inst:UoAO}} \and
	Michael Bonnefoy \inst{\ref{inst:IPAG}} \and 
	Faustine Cantalloube \inst{\ref{inst:LAM}}\and
	H. Jens Hoeijmakers \inst{\ref{inst:sweden}} \and 
	Ulli Käufl \inst{\ref{inst:ESO}} \and
	Markus Kasper \inst{\ref{inst:ESO}} \and
	Anne-Lise Maire \inst{\ref{inst:IPAG}}\and
    Mathilde Mâlin \inst{\ref{inst:LESIA}} \and
    Ralf Siebenmorgen \inst{\ref{inst:ESO}} \and
	Ignas Snellen \inst{\ref{inst:Leiden}} \and
	G\'{e}rard Zins \inst{\ref{inst:ESO}} 
	}


   \date{Received 10/05/2022; accepted 08/03/2023}

  \abstract
   {$\beta$ Pictoris is a young nearby system hosting a well-resolved edge-on debris disk, along with at least two exoplanets. It offers key opportunities for carrying out detailed studies of 
    the evolution of young planetary systems  and their shaping soon after the end of the planetary formation phase.}
   {
   We analyzed high-contrast coronagraphic images of this system, obtained in the mid-infrared, taking advantage of 
   the NEAR experiment using the VLT/VISIR instrument, which provides access to adaptive optics, as well as phase coronagraphy.
The goal of our analysis is to investigate both the detection of the planet $\beta$ Pictoris b and of the disk features at mid-IR wavelengths.
In addition, by combining several epochs of observation, we expect to constrain the position of the known clumps and improve our knowledge on the dynamics of the disk.
   }
   {
   We observed the $\beta$ Pictoris system over two nights in December 2019 in the $10-12.5\, \muup$m coronagraphic filter. To evaluate the planet b flux contribution, we extracted the photometry at the expected position of the planet and compared it to the flux published in the literature as well as the one we measured with SPHERE spectroscopy in the near-IR. In addition, we used previous data from T-ReCS and VISIR in the mid-IR, updating the star's distance, to study the evolution of the position of the southwest clump that was initially observed in the planetary disk back in 2003.
   }
   {
   While we did not detect the planet b, we were able to put constraints on the presence of circumplanetary material,
   ruling out the equivalent of a Saturn-like planetary ring around the planet. The disk presents several noticeable structures, including the known southwest clump. Using a 16-year baseline, sampled with five epochs of observations, we were able to examine the evolution of the clump. We found that the clump orbits in a Keplerian motion with a semi-major axis of 56.1$^{+0.4}_{-0.3}$\,au.
   In addition to the known clump, the images clearly show the presence of a second clump on the northeast side of the disk as well as possibly fainter and closer structures that are yet to be confirmed. Furthermore, we found correlations between the CO clumps detected with ALMA and the northeastern and southwestern clumps in the mid-IR images.
   }
   {
   If the circumplanetary material were located at the Roche radius, 
   the maximum amount of dust determined from the flux upper limit around $\beta$ Pictoris b would correspond to the mass of an asteroid of 5\,km in diameter. Finally, the Keplerian motion of the southwestern clump is  possibly indicative of a yet-to-be detected planet or signals the presence of a vortex.
   }

   \keywords{Exoplanets, Direct Imaging, High Angular Resolution}

\authorrunning{Skaf et al.}
\maketitle

\section{Introduction} \label{sec:intro}

Exploring the diversity of planetary systems requires a broad spectral range tuned for detecting various components in these environments: gas, dust, planetesimals, and planets.
The past two decades have seen the development 
of extreme adaptive optics (AO)  on monolithic telescopes, a technology 
that
allows 
for  high angular resolution and high contrast imaging to be performed at both visible and near-infrared (NIR) wavelengths. 
Instruments such as the Spectro-Polarimetric High-contrast Exoplanet REsearch (SPHERE, \citealt{Beuzit2019}) are sensitive to the thermal emission of young giant planets, as well as the scattered light from dust particles. 
At the same time, the Atacama Large Millimeter/submillimeter Array (ALMA) has revolutionized observations of the thermal radiation emitted from gas components and millimeter dust grains in both debris disks and gas-rich protoplanetary disks \citep[e.g.,][]{ALMA2015}.
The intermediate spectral range, the mid-IR, has received much less attention thus far, primarily because of the high thermal background contamination in ground-based observations. However, this paradigm is about to change thanks to the James Webb Space Telescope and its Mid-Infrared Instrument (MIRI, \citealt{Rieke2015, Wright2015} and references there-in), as well as (later in the decade) the upcoming Mid-infrared ELT Imager and Spectrograph (METIS, \citealt{Brandl2018}).

In the meantime, the New Earths in the $\alpha$ Centauri Region project \citep[NEAR,][]{Kasper2019}  was installed at the VLT to perform a deep observing campaign of $\alpha$ Centauri \citep{Wagner2021} and (as  part of the Breakthrough Initiatives Foundation) offered an upgrade of the existing VLT spectrometer and imager for the mid-infrared instrument (VISIR, \citealt{VISIR_Lagage}), which operates in the L, N, and Q bands.
Then, VISIR was moved to UT4 for a period of time to benefit from the Adaptive Optics Facility on the secondary mirror \citep{Arsenault2017} and an annular groove phase mask (AGPM) coronagraph \citep{MawetAGPM2005,Delacroix2012} was installed, operating in a single broad filter covering from 10\,$\muup$m to 12.5\,$\muup$m, with the filter centered at 11.25\,$\muup$m \citep{Maire2020}. We took advantage of the NEAR Science Demonstration program to observe the $\beta$ Pictoris system.

$\beta$ Pictoris is a young  ($\sim18.5^{+2.0}_{-2.0}$ Myr, \citealt{Miret-Roig2020}) and bright main sequence pulsating star. Since the discovery of its debris disk by \citet{SnithTerrile1984}, this system has been extensively observed to search for signs of planets. First indirect clues of a planet were brought by the detection of star-grazing comets falling onto the star \citep{Lagrange1988} and of a warp in the disk attributed to the gravitational interaction between an unseen massive body (planet, brown dwarf, etc.) on an inclined orbit and the planetesimals in the disk \citep{Mouillet1997}.
According to surface brightness measurements, the planetesimals are distributed within $\sim$120\,au, while the inner part inside $\sim$80\,au is warped \citep{Augereau2001}. In scattered light, small dust particles, blown by the stellar radiation pressure, have been observed at even larger projected separations \citep{Janson2021}.

Years later, observations with AO of the NaCo instrument at the VLT led to the detection of the thermal emission from the giant planet $\beta$ Pictoris b \citep{Lagrange2008, Lagrange2010}.
Finally, in 2019, a second planet $\beta$ Pictoris c, was identified via radial velocity as a result of nearly ten years of monitoring the host-star with the High Accuracy Radial velocity Planet Searcher (HARPS) instrument \citep{2019L_bpic}. This planet c was then confirmed with long-baseline optical interferometry \citep{nowak_m_direct_2020, lagrange_unveiling_2020}.

The $\beta$ Pictoris system is certainly unique to probe the 
mechanisms of planetary formation. A large fraction of the previous studies focused on the characterization of the disk and, in particular, on the possible interaction between planets with the dust and gas components by identifying the different features in various spectral regimes and spatial scales. In that respect, \citet{Apai2015} provided a comprehensive analysis of the system by using data obtained in various spectral bands: optical, near IR, mid-IR, and sub-millimeter. Asymmetries, especially between the northeast and the southwest arms of the edge-on disk, are reported at all wavelengths both in scattered light, tracing the submicron-size grains, and in emission, through the analysis of
micron-size and sub-millimeter-size grains. A striking feature in the mid-IR is the clump identified in the southwest by \cite{Telesco2005}, located at a projected separation of 52\,au. 
The clump was confirmed by \citet{Pantin2005}, and recovered by \citet{danli2012} with a small offset, which was interpreted as orbital Keplerian motion. In addition, this asymmetric feature appears at $\sim10\,\muup$m and fades away at $\sim20\,\muup$m, indicative that its flux is dependent on either temperature, grain size, composition, or a combination of these three effects. \cite{Telesco2005} argued that
this clump could be the result of either collisions among planetesimals trapped in resonance, or the dismantling of a more massive object. Finally, \cite{Okamoto2004} reported amorphous silicates peaks at about 6, 16, and 30\,au of radial distance from the central star, and a recent study revisited {\it Spitzer} data identifying various rings based on amorphous silicates peaks \citep{Lu_2022}.

In addition to the dust clumps observed in the mid-IR, ALMA identified 
the presence of spatially extended features from both a CO \citep{Dent2014, Matra2017} and neutral carbon \citep{Cataldi2018} gas. 
The CO clump can be readily explained because CO would be mainly produced in the brightest dust clump and is overabundant there. Indeed, CO would not have time to spread around its orbit and become axisymmetric because the photodissociation timescale is much smaller than the orbital timescale and CO cannot complete a full orbit. Thus, when observed edge-on, it still looks like a clump similar to the dust clump. 
With regard to carbon, a clump is much more complicated to explain with current models { as carbon gas is expected to survive over Myr timescales \citep{Kral2016} and to quickly become symmetrical in azimuth because of collisions with ambient gas (even if it were produced in an asymmetric clump initially).}

In this work, we analyze, for the very first time, the mid-IR images of $\beta$ Pictoris obtained using adaptive optics technology, which enables us to reach an unprecedented angular resolution and contrast at this wavelength for this emblematic system. Furthermore, we revisit former VISIR data to perform a multi-epoch comparison. 
As a result, we provide important constraints on the planet and the disk structures.

The paper is organized as follows: Sect. \ref{sec:obs} presents the context of the observations, as well as the data reduction and post-processing. We discuss the non-detection of the planet b in Sect. \ref{sec:planetflux} and we put some constraints on the presence of dust material around the planet in Sect. \ref{sec:planetdust}. In Sect. \ref{sec:disk}, we analyze the disk structures and measure the variations of the southwest clump position to better constrain its orbital radius in the system, which is described in Sect. \ref{sec:disk_astrometryclump}. Section \ref{sec:disk_otherclumps} covers the other clumps detected in the disk. 
Finally, we discuss the implications of these observations in Sect. \ref{sec:discussion} and we present our conclusions in Sect. \ref{sec:conclusion}.

\section{Observations and data reduction}
\label{sec:obs}

\begin{table*}[t]
\centering
\begin{tabular}{llllll}
\hline
                    &  \textbf{2003-12-30} & \textbf{2004-08-30} &\textbf{2010-12-16} & \textbf{2015-09-01} & \textbf{2019-12-15} \\ 
\hline
\hline
\textbf{Instrument} &  T-ReCS        &  VISIR     & T-ReCS        & VISIR         & NEAR          \\ 
\textbf{Telescope}  & Gemini   & VLT UT3    & Gemini        & VLT UT3       & VLT UT4       \\ 
\textbf{Filter}     &  12.3\,$\muup$m   & 11.7\,$\muup$m       & 10\,$\muup$m     & 11.7\,$\muup$m       & 11.25\,$\muup$m        \\ 
\textbf{AO+Coronagraph}        & No             & No            & No            & No            & Yes           \\
\textbf{Reference}      & \cite{Telesco2005}             & This study            & \cite{danli2012}            & This study            & This study           \\
\hline
\textbf{C1 pos (au)} & $52.9^{+0.5}_{-0.5}$ & $52.7^{+0.6}_{-0.6}$ & $55^{+0.5}_{-0.5}$ & $55.5^{+0.4}_{-0.4}$ & $56.1^{+0.3}_{-0.3}$ \\ \hline
\end{tabular}
\caption{Summary of the data used and the observing modes with different instruments. The last line presents the evolution of the C1 position over the years, considering a stellar distance of 19.63 pc.}
\label{tab:datatable}
\end{table*}

\subsection{VISIR NEAR science demonstration data}

\subsubsection{Project description}

The VLT mid-infrared imager, VISIR, has been modified in the framework of the NEAR project \citep{Kasper2019} for an observation campaign of the $\alpha$ Centauri system aiming at detecting 
{terrestrial} planets. NEAR is
supported by ESO and the Breakthrough Initiatives Foundation.
The project was designed to push VISIR N-band performances to their limit, in terms of sensitivity and contrast.
To that aim, VISIR was equipped with phase mask vortex coronagraphs (AGPM) \citep{MawetAGPM2005, Delacroix2012} optimized for the  $10-12.5\,\muup$m band.
The control of the centering of the star into the coronagraph was performed with the quadrant analysis of coronagraphic images for tip-tilt sensing estimator (QACITS, \cite{Huby2015, Maire2020}).
Also, NEAR has been coupled to a visible wavefront sensor that controls the deformable secondary mirror (DSM) of the adaptive optics facility (UT4), 
allowing access to extreme adaptive optics regime and reaching Strehl ratios higher than 90\% at 12 $\muup$m.

\subsubsection{Observing setup}

$\beta$ Pictoris was observed with NEAR on the 12th and 
the 15th of December 2019 (program ID 60.A.9107 (K), P.I. NEAR team).
Coupling NEAR with the DSM of the VLT allowed for the use of AO without increasing the number of warm optics that would add to the thermal background. The NEAR spectral filter transmits light from 10 to 12.5 $\muup$m and the full width at half maximum (FWHM) is $\sim$ 0.28$\arcsec$, or $\sim$ 6 pixels. The Cassegrain instrument was fixed in pupil-stabilized mode during the observations. 
To suppress the large thermal background, a chopping frequency of 8.33\,Hz has been adopted 
and a chop throw of 4.5$''$. This relatively high frequency allowed to significantly 
decrease the low-frequency noise excess of the {\it Aquarius} type detector \citep{dives2014} and reach
background limited sensitivity performances. To compensate for the difference in optical path induced by the chopping, the nodding was performed by slightly slewing the telescope every minute. The subtraction of the chopping yielded to have
two images of the star taken at different positions: one coronagraphic and the other an off-axis, non-coronagraphic image. As a result, there are two negative images of the star on the side of the central coronagraphic image.

The detector integration time was set to 6\,ms to avoid saturation in the background. Ten frames were acquired and stacked in each DSM chopper position, out of which the  first two were discarded to suppress any PSF smearing due to the DSM settling time. The recorded data have thus a time frame of 160 ms.
Simple chopping leaves some low spatial frequency residuals due to inhomogeneities of the thermal footprint.
Nodding was thus also applied to calibrate these residuals. The nodding offset was parallel and of the
same amplitude as the chopping so that the target was always kept within the detector's field of view. A temporal binning of chopped-nodded frames was performed 
to produce a data cube in which each frame corresponds to an integration time of about 5s.
The data from
the 12th and the 15th were combined
corresponding to 2442\,s and 3180\,s on source respectively. During these observations, the seeing ranges were $0.7-0.9''$ and $0.55-0.80''$ on the 12th and the 15th respectively, and the coherence time was $\sim$5\,ms and $\sim$4\,ms, respectively. The precipitable water vapor (PWV) was $\sim$5\,mm and $\sim$3\,mm respectively, as well.

\subsubsection{Post-processing}

\begin{figure*}[t]
\centering
\includegraphics[width=16cm]{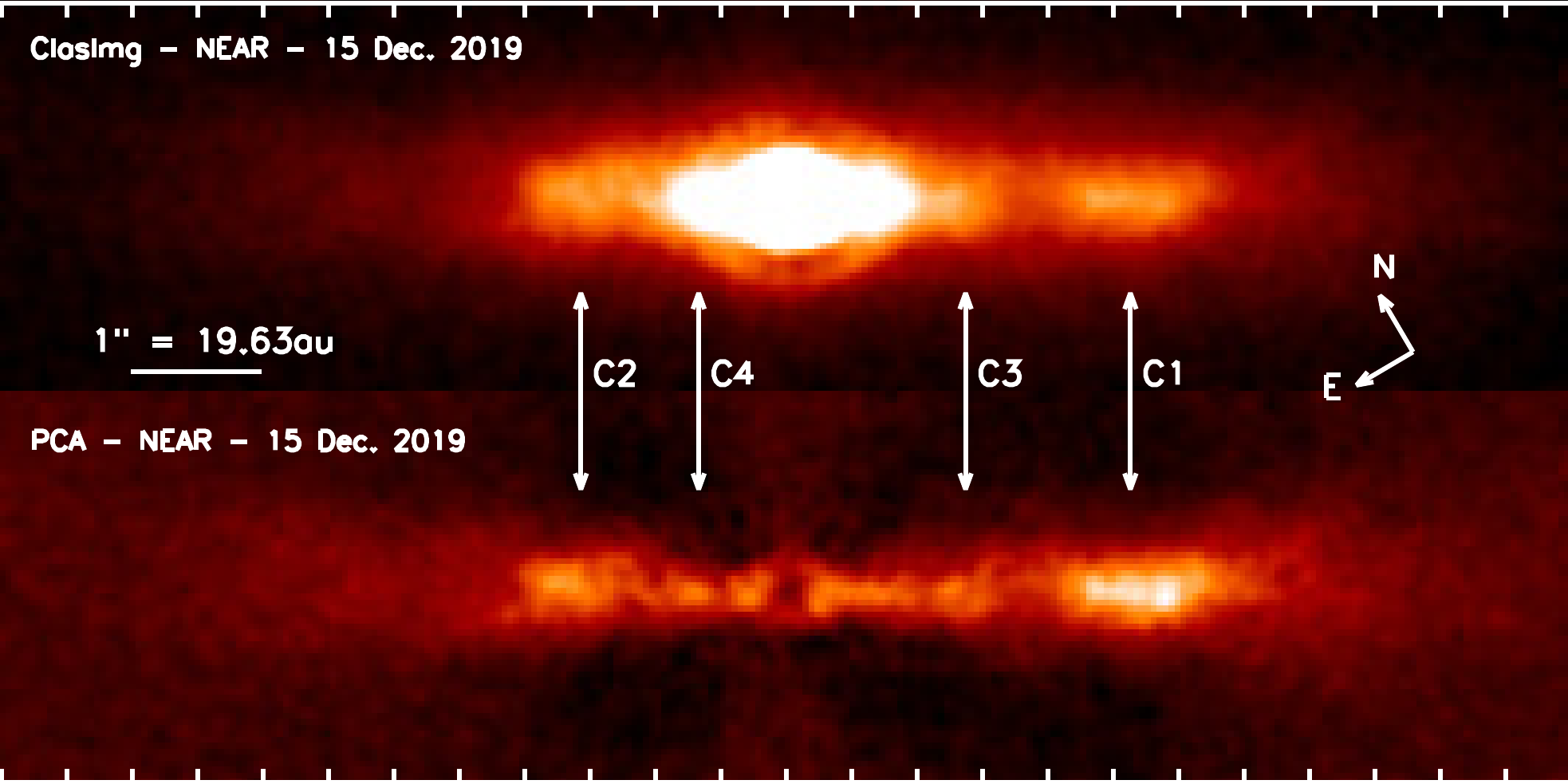}
\caption{%
Image of the disk at 12 $\muup$m, using a PCA reduction (bottom), and a ClasImg reduction 
performed with SpeCal. The disk has been rotated 
by
30 degrees in those images with respect to due north. 
Labels indicate the dust clumps described in Sect. \ref{sec:disk}.
}
\label{fig:PCAnoADI}
\end{figure*}

After the observations, frame selection was performed on the data cubes, based on the criteria of residual flux behind the coronagraphic mask. The 10\% worst frames for which the star was slightly off-centered were rejected because otherwise resulting in poor stellar subtraction during the next step.  
After frame selection, the remaining on-source telescope time (behind the mask) is about 5060\,s. We took advantage of the SPHERE Calibration Tool, SpeCal \citep{GalicherSpecal2018}, which can be easily adapted to any instrument by providing minor modifications of the input data to match the SpeCal format.
SpeCal has been developed to accurately detect the faintest point sources in high-contrast imaging data, offering several types of algorithms for data analysis, such as angular differential imaging \citep[ADI,][]{2006_Marois_ADI} combined with
principal component analysis \citep[PCA,][]{Soummer2012} or TLOCI \citep{2007_loci_lafreniere}. The simplest processing provides a direct stacking of frames once compensated by the field rotation. {This is the classical averaging 
described in \citet[][ClasImg, for classical imaging]{GalicherSpecal2018}.}
Figure \ref{fig:PCAnoADI} presents the resulting images, after a ClasImg and PCA reduction.

\subsection{VISIR archival data}

Non-coronagraphic observations of $\beta$ Pictoris were obtained with VISIR on 2004 -- program ID 60.A-9234(A) and 2015 (program ID 95.C-0425(A) -- with the VLT. The observations were performed in a standard chopping-nodding mode. 
The data were reduced by combining all the chop-nodded frames into a single one. Each observation was typically one hour long.
The data were consequently flux calibrated by observing with the same settings an infrared standard (HD\,42540, spectral type K2/3III, J=2.90, H=2.24, K=2.09) just before the $\beta$ Pictoris observations.  
The reduced data are further examined in Sect. \ref{sec:disk_astrometryclump} to perform a comparison with the NEAR observations.

\section{Searching for $\beta$ Pictoris b}
\label{sec:planetflux}

\subsection{A non-detection at mid-infrared wavelengths}
\label{sec:nondetection}

Using various SpeCal's algorithms, we performed several tests to reduce the stellar contribution and to reveal the expected planet. We compared the contrast curves provided by the cADI, ClasImg, PCA, and TLOCI algorithms: we found that the PCA reduction provides the deepest contrast curves at the expected separation of the planet.
All the assessments provided a non-detection outcome, preventing us from identifying the giant planet $\beta$ Pictoris b in the NEAR data.
Observations obtained at a similar epoch with SPHERE yielded an angular separation of $301\pm 4$\,mas with respect to the star \citep{lagrange_unveiling_2020}. At such a distance, which corresponds to $1\times\lambda/D$ at this wavelength ($\lambda=11.25\,\muup$m), the AGPM coronagraph would transmit only $32\%$ of the flux of a point-source (see Figure \ref{fig:transmission}). Taking this parameter into account, we further explored the sensitivity to point sources by injecting fake planets built from the non-coronagraphic star's image at various contrasts, and processed with a set of ADI algorithms. We concluded that the PCA algorithm is providing the best contrast with ten modes removed.

One of the SpeCal output provides an estimation of the azimuthal contrast level
for each angular distance. Therefore, we estimated the limit of detection in the PCA image. We note that the disk itself and the background are 
the major contributions to the noise, far beyond the speckle noise. 
The transmission curve of the coronagraph was 
computed by comparing the maximum intensity of the coronagraphic image at an increasing separation from the center, normalized to the PSF (non-coronagraphic) maximum intensity (Figure \ref{fig:transmission} in the appendix).
Figure \ref{fig:cc} shows  the 5$\sigma$ contrast curve corrected from the coronagraphic transmission. The contrast at the expected position of the planet $\beta$ Pic b at 0.3" is $5.10^{-3}$.

\begin{figure}[t]
\centering
\includegraphics[width=9cm, trim=0.5cm 0cm 2cm 0cm, clip]{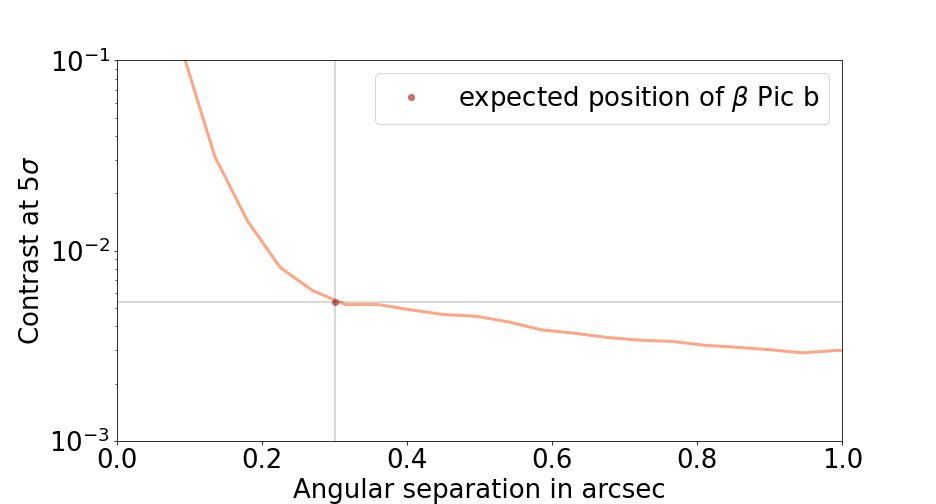}
\caption{Contrast curve at 5$\sigma$ as measured along the disk spine across a width of 1", as measured on the PCA image. The contrast of $\beta$ Pic b with respect to the star and its distance to the star 
is indicated by the grey lines. 
}
\label{fig:cc}
\end{figure}

\subsection{Upper-limit constraints on the spectral energy distribution}

The spectral energy distribution of $\beta$ Pic b has been studied in a number of papers \citep{Bonnefoy2011, Bonnefoy2013, Currie2013,Bonnefoy2014, Morzinski2015, exorem2015, Chilcote2017,NowakGravity2020}. 
The purpose of this section is not to carry out a similarly
detailed analysis, but instead to perform an order of magnitude comparison between the detection limit presented in Sect. \ref{sec:nondetection}, by using atmospheric models predictions, determined from the planet's flux in the near-IR.

\begin{figure}[t]
\centering
\includegraphics[width=9cm, trim=0cm 0cm 2cm 0cm, clip]{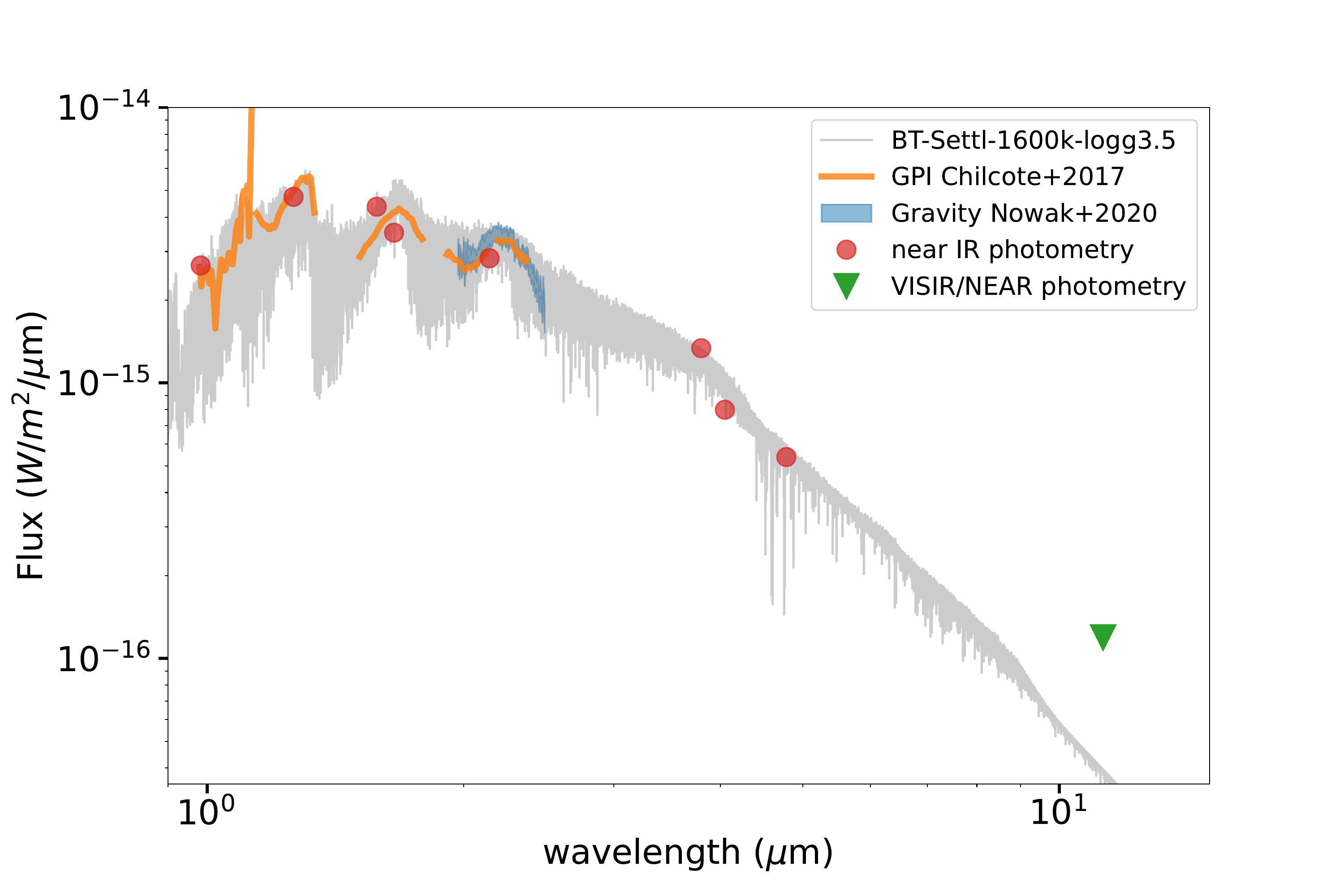}
\caption{Spectra of $\beta$ Pictoris b obtained by fitting forward BT-Settle models to data obtained with several instruments (color-coded in the legend). The upper flux limit of the NEAR data is marked by a green triangle. }
\label{fig:spectrum}
\end{figure}

First of all, we calculated the stellar flux density $F_{\lambda}$ by assuming a BT-NEXTGEN\footnote{\url{https://phoenix.ens-lyon.fr/Grids/BT-NextGen/}} model at T$_{\rm eff}$ = 8000\,K, $log\,g=4.0$, with solar metallicity, as in \citet{Chilcote2017}, together with a star's radius of $1.65\,R_{\odot}$ and a distance of 19.6\,pc (EDR3, \citealt{Gaia2021}). Further, this stellar model was normalized to match the actual photometry of $\beta$ Pic A \citep{Bonnefoy2013}. 
As for the planet spectral energy distribution, we used broad bands and narrow bands data (Y to M) from \citet{Bonnefoy2013}, as well as the Gemini Planet Imager (GPI) spectrum presented in \citet{Chilcote2017} renormalized to the estimation of the star's flux density from \cite{Bonnefoy2013}.
These photometric data are displayed in Figure \ref{fig:spectrum}.

The upper limit of the planet contrast measured at 11.25\,$\muup$m (the NEAR wavelength) translates to a flux density of $1.146\times10^{-16}$\,W\,m$^{-2}\,\muup$m$^{-1}$. A straightforward comparison with a BT-Settl model of the planet at T$_{eq}$=1600\,K and $\log{g}=3.5$ unambiguously shows that the non-detection with NEAR does not allow to put meaningful constraints on the planetary atmospheric properties, the upper limit of the observed flux density being about three times larger than the model's expectation from the near-IR detection (Figure \ref{fig:spectrum}).

\subsection{Exploring the presence of circumplanetary material}

\label{sec:planetdust}
\subsubsection{Context}

We often assume young giant planets 
to often be surrounded by a circumplanetary disk that later
accretes onto the planet and produces satellites. 
The satellite-forming process is expected to leave a gas-free dusty disk, consisting of ring structures filling the planet's Hill sphere, which will eventually disperse. A 18\,Myr-old planet like $\beta$ Pic b could still be surrounded by such a fading, and now optically thin circumplanetary disk. 
The planet's Hill sphere has been indicated as a potential explanation for a $\sim4$\% photometric variation observed in 1981 \citep{lecavelier95}. 
However, continuous monitoring of the expected 2017 and 2018 Hill sphere stellar transits was  not able to detect flux variation due to circumplanetary material \citep{Kenworthy2021}; instead, it could place an upper limit of $\sim1.8\times10^{19}$kg of dust in the planetary Hill sphere. We aim to investigate to what extent our 12\,$\muup$m non-detection could place an upper limit on the amount of circumplanetary material. 
We recall that the angular resolution of NEAR at $11.25\,\muup$m corresponds to a patch of  $5-6$\,au, which is much larger than the Hill radius of the planet (about 1.2\,au for $\beta$ Pic b), so that any circumplanetary material will appear unresolved. 

This prevents us from constraining any further the dust's spatial location and its size distribution around the planet.
Consequently, by considering simple but physically relevant assumptions on the dust location and grain size distribution, we investigate the possible order of magnitude of the total dust mass that agrees with the flux upper limit measured at the planetary location.

\subsubsection{Model}
Schematically, at a given wavelength $\lambda$, dust grains are efficient emitters or scatters provided they have a radius, $s_{\rm dust}$, that is 
on the order of $s_{\lambda} \sim \lambda/2\pi $. For the NEAR filter, we 
calculated
$s_{\lambda}=1.9\,\muup$m, which is a value
relatively close to the minimum size, $s_{\rm blow}$, below which grains are blown out of the system by stellar radiation pressure. In fact, if we assume that grains are produced from progenitors on a circular orbit, then such grains will be ejected if the ratio $\beta$ between radiation pressure and stellar gravity is higher than 0.5. Taking the standard expression for $\beta$ given by \citet{Krivov2010}:
\begin{equation}
    \beta \approx \frac{0.57}{s_{\rm blow} [\muup \rm m]}\frac{1}{\rho_{\rm dust}  [\rm g\,\rm cm^{-3}]}\frac{L_{\star}}{L_{\odot}}\frac{M_{\odot}}{M_{\star}},
\end{equation}
with $M_{\star}$ and $L_{\star}$ being the mass and luminosity of the central star,
and assuming $\rho_{\rm dust}=2.7$ g$\cdot$cm $^{-3}$, typical astrosillicates density, $L_{\star}=8.7$\,L$_{\sun}$ \citep{Crifo1997} and $M_{\star}=1.77$\,M$_{\sun}$ \citep{lagrange_unveiling_2020}, we get  
$s_{\rm blow} \approx 2.1\,\muup$m. 


By making the 
assumption that grains are produced by a collisional cascade starting from larger planetesimal-like bodies (be it in a circumstellar cloud or disc), then we can assume that they follow a standard size distribution in $dn = s^{-3.5} ds$ down to $s=s_{\rm blow}$\footnote{For the order of magnitude  related to our estimates, we ignore the fact that size distributions can depart from this standard behavior of real systems \citep{theb07}. We also note that our argument about the smallest grains dominating the total cross-section is valid for any size distribution $dn = s^{-q} ds$ or index $q>3$.}. 
For such a distribution, the geometrical cross-section is dominated by the smallest grains just above $s_{\rm blow}$ \citep{theb16}. Since $s_{\lambda} \sim s_{\rm blow}$, this also means that the flux at $\lambda$ is dominated by the smallest grains in the size distribution. For the sake of simplicity, we thus consider here a single-sized dust distribution made of $s_{\rm dust}=2\,\muup$m grains.

\begin{figure}[t]
    \centering
    \includegraphics[trim=1cm 0.cm 2cm 2cm,clip,width=0.5\textwidth]{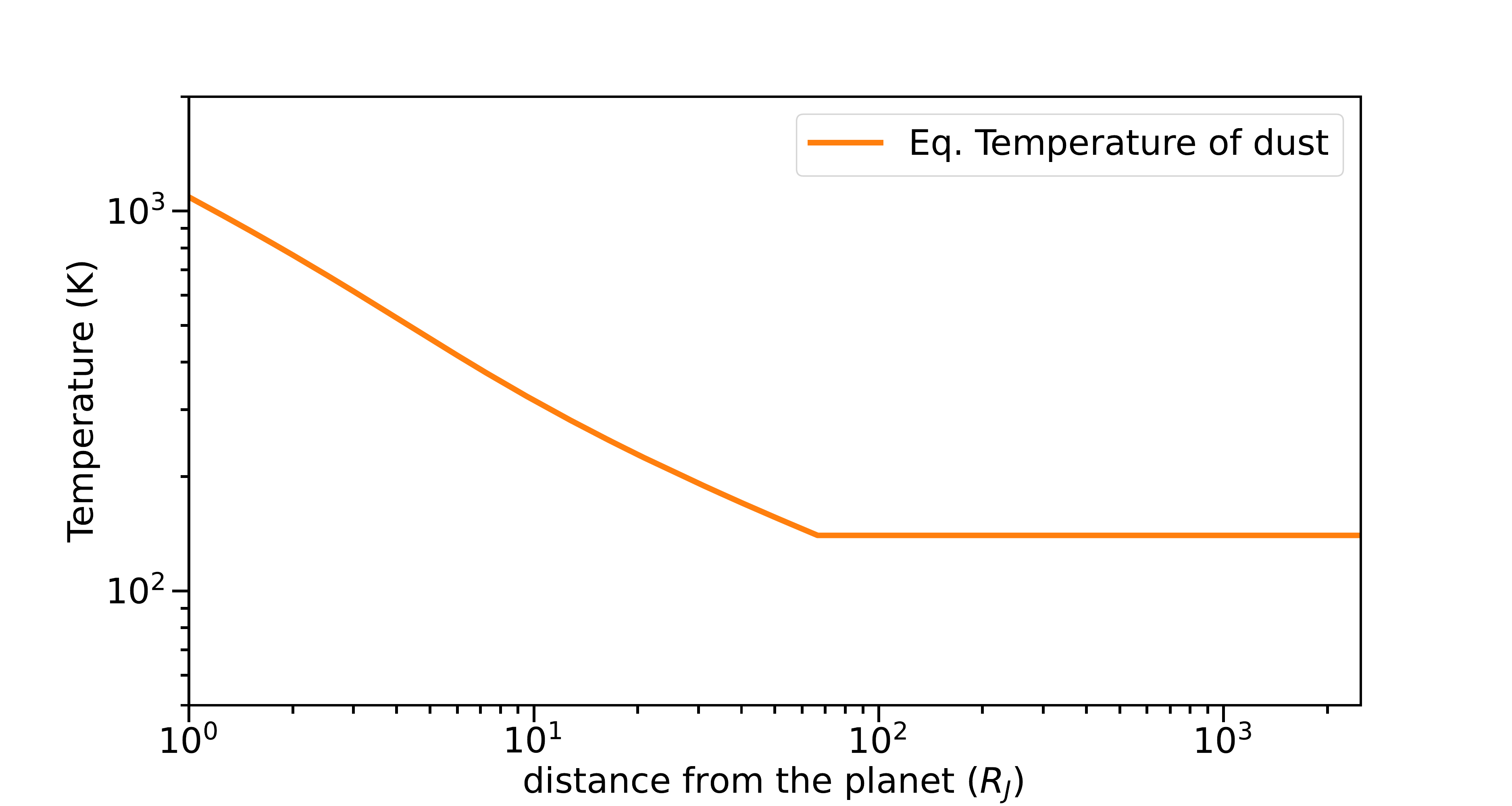}
    \caption{
    Dust equilibrium temperature, calculated with Mie theory, of a 2\,$\muup$m grain particle located within a radius of $\sim 100\, R_J$ from $\beta$ Pictoris b and assumed to be constant further out up to $R_{\rm Hill}$.
    }
    \label{fig:tempdust}
\end{figure}

To estimate the temperature $T_{\rm dust}$ and flux $F(T_{\rm dust})$ emitted by each dust grain, we consider the framework of the Mie theory for compact astrosilicate grains and use the GRaTer radiative transfer code \citep{Augereau1999, Olofsson2020} to estimate the absorption coefficient, $Q_{\rm abs}$. We note that we here have two heating sources for the grains: the radiation from the star and the radiation from the 
warm
young planet. Given the estimated temperature of the planet \citep[$\sim1700$\,K,][]{Bonnefoy2013}, the stellar parameters of $\beta$ Pictoris, and the separation of 9.8\,au between the host star and the planet, 
we find that grain heating by the planet dominates up to a distance $\sim 100$ R$_J$ from the planet and that the stellar 
heating dominates beyond that. At the tipping point between these two domains, the grain temperature is of the order of $T_{\rm dust} \sim 140$\,K (Figure \ref{fig:tempdust}).



We now assume that a grain emits as a black body weighted by $Q_{\rm abs}$, according to the temperature previously determined. Since the dust scattering at mid-IR is negligible against the emission, the flux re-radiated by a dust particle is as follows:
\begin{equation}
F(T_{\rm dust},\lambda) = Q_{\rm abs}\frac{2\pi hc}{\lambda^5}\frac{1}{e^{hc/kT_{\rm dust}\lambda}} \frac{4\pi s_{\rm dust}^2}{4\pi d_{\rm pd}^2}
,\end{equation}

with $d_{\rm pd}$ being the distance planet to dust, $c$ being the speed of light, $h$ being the Planck constant, and $k$ being the Boltzmann constant.

Considering the flux upper limit on the planet position (derived in Sect. \ref{sec:planetflux}), 
 $F_{\rm limdet}=1.146\times10^{-16}$\,W\,m$^{-2}\,\muup$m$^{-1}$
and in the optically-thin hypothesis, we then derive, at each radial distance from the planet, the maximum number of 2\,$\muup$m-size dust particles as $N_{\rm dust}=F_{\rm limdet} / F(T_{\rm dust},\lambda)$.

For spherical grains, the corresponding total mass of dust M$_{\rm dust}$ is then:
\begin{equation}
M_{\rm dust} = N_{\rm dust}\,\rho_{\rm dust}\frac{4}{3}\pi s_{\rm dust}^3
,\end{equation}
which we can convert to an equivalent radius of a spherical parent body, R$_{\rm parent}$, of the same density:
\begin{equation}
R_{\rm parent} = \left(\frac{M_{\rm dust}}{\rho_{\rm dust}}\frac{3}{4 \pi}\right)^{1/3}
.\end{equation}

\begin{figure}[h]
\centering
\includegraphics[width=9cm, trim=0cm 0cm 0cm 0cm, clip]{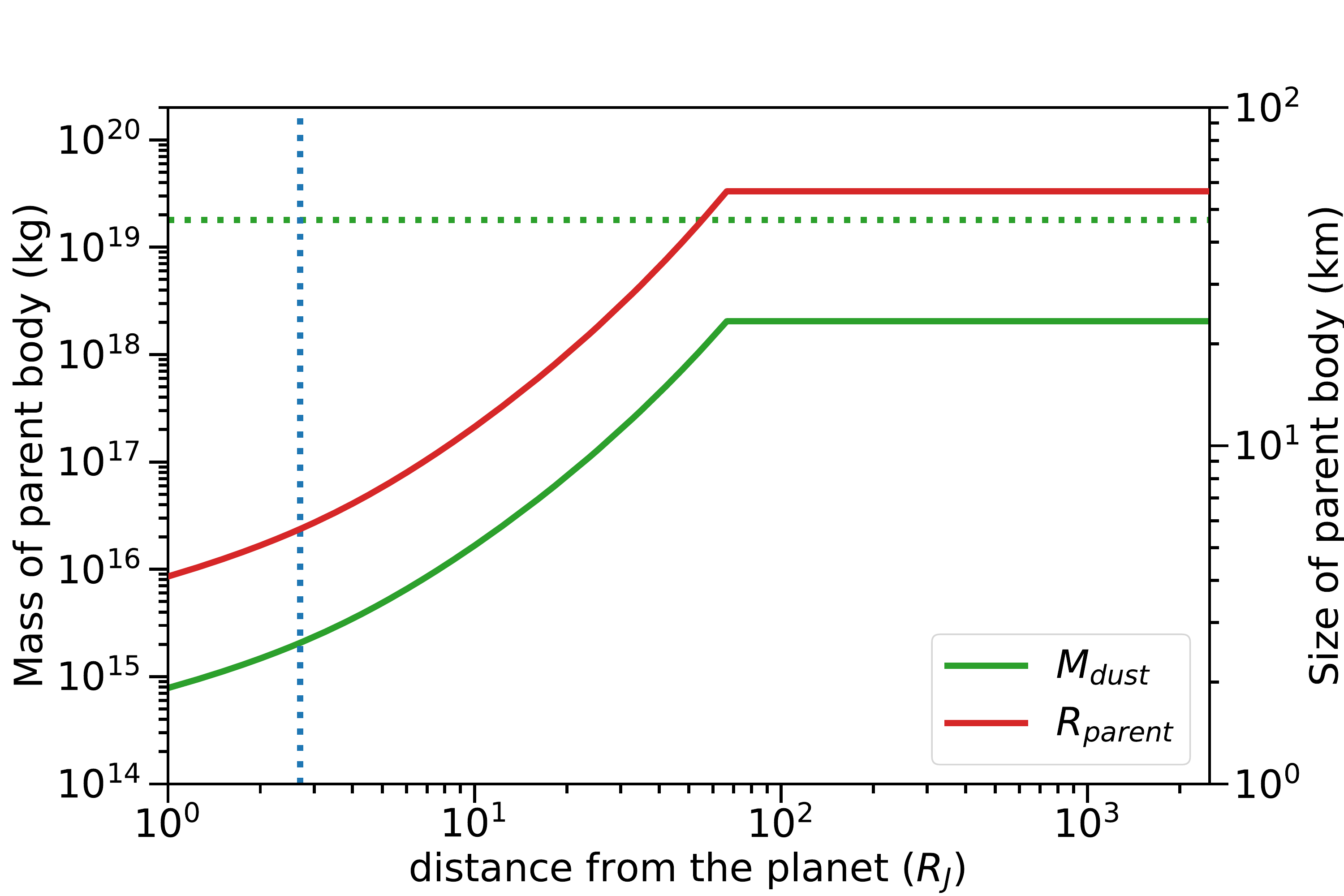}
\caption{Mass (green line) and size (red line) of a parent body able to produce by collision a cloud of dust of which the emitted flux in the NEAR bandpass could be compatible with $F_{\rm limdet}$, at the planet position. The green dotted line represents the upper limit of \citet{Kenworthy2021}. The blue dotted line stands for the Roche limit. 
}
\label{fig:massvsdist}
\end{figure}

We summarize our results in Figure \ref{fig:massvsdist}, which plots $M_{\rm dust}$ and $R_{\rm parent}$ as a function of radial distance to the planet up to the Hill sphere limit $R_{\rm Hill}$ at $\sim1.2$\,au,  defined as:
\begin{equation}
R_{\rm Hill} = a_{\rm planet}\times\left(\frac{M_{\rm planet}}{3M_{\star}}\right)^{1/3}
\label{eq:hillequation}
,\end{equation}

\noindent with $a_{\rm planet}$ as the planet's semi-major axis, $M_{\rm planet}$ as the planet's mass, and $M_{\star}$ as the mass of the central star.

We see that within the 
100\,R$_J$ radius domain, where heating is dominated by the planet, the dust mass required to emit $F_{\rm limdet}$ increases with increasing distance from the planet. 
Such a result was expected due to the dust temperature decrease observed within this domain (Figure \ref{fig:tempdust}).
On the contrary, in the region beyond $\sim100\,R_J$, where stellar heating dominates, both $M_{\rm dust}$ and $R_{\rm parent}$ stay constant. This means that, within the simplified frame of our model, it is impossible to distinguish between dust located at $100\,R_J\sim 7.15\times10^6$\,km and $R_{\rm Hill}\sim 1.78
\times10^8$\,km. 
Interestingly, the value of $M_{\rm dust}\sim5\times10^{18}$\,kg that we derived in the stellar radiation-dominated domain is relatively close to the upper limit of $1.8\times10^{19}$\,kg derived by \citet{Kenworthy2021} during the hypothetical transit of the Hill sphere of the planet in 2017-2018 (displayed as a dotted green line in Figure \ref{fig:massvsdist}).

We note, however, that we obtained different dust mass constraints in the more compact planet-heating dominated domain. For example, if we consider the case of a planetary ring extending out to the Roche radius at $\sim 2.7\,R_J $, then the maximum dust mass value would be only $\sim\num{2e15}$\,kg, corresponding to the mass of a $\sim5\,$km-sized object. This is $\sim700$ times less than the estimated mass in Saturn's ring \citep{Iess2019}, so we can 
rule out the presence of a massive Saturn-like planetary ring around $\beta$ Pictoris b.

\section{Dissecting the disk morphology}
\label{sec:disk}

In this section, we present and analyze the morphology of the disk structures as observed at mid-IR wavelengths with NEAR-VISIR. The $\beta$ Pictoris disk is seen edge-on and extends to a projected separation of $\sim 5''$ that is equivalent to about 100\,au. Figure \ref{fig:PCAnoADI} presents a cropped, rotated image of the disk, corresponding to the 
2019 observations with NEAR, for two different reduction methods. The ClasImg reduction (described in Sect. \ref{sec:nondetection}) performs a derotation and averaging of the frames, while PCA provides better rejection of the starlight, but comes with self-subtraction artifacts \citep{Milli2012}. The ClasImg image has the advantage to preserve the disk photometry and morphology, but the star's diffraction pattern dominates at short separations ($<0.75''$). The disk image in the thermal regime features notable structures in the form of several clump-like patterns, as indicated in Figure \ref{fig:PCAnoADI}. In particular, 
the southwest (SW) clump, labeled C1, discovered
in previous studies \citep{Telesco2005, danli2012}, and located at a separation of $\sim 2.8''$ (projected distance of $\sim$55\,au), is the most obvious feature in the image. Another clump (C2) is visible in the northeast (NE) side at a position of $\sim1.7''$ (projected distance of $\sim$33\,au). C1 and C2 are both rather broad with a full width at half maximum of $\sim1.3''$ and $\sim0.8''$, respectively. 
Although they were only marginally detected, we also identified 
two more clumps: C3 at the SW (1.5$''$ equivalent to a projected separation of $\sim$30\,au) and C4 in the NE (0.8$''$ equivalent to a projected separation of $\sim$16\,au that is less than $3\lambda/D$). 
Still, the reliability of C4 as a real disk structure may need further confirmation to disentangle from diffraction residuals.
Contrary to near IR observations, the NEAR image does not reveal any sign of the 
warp \citep{Mouillet1997}, a feature that is only observed in scattered light.

\subsection{Orientation of the disk spine}
\label{sec:disk_pa}

\begin{figure}[t]
\centering
\includegraphics[width=0.54\textwidth]{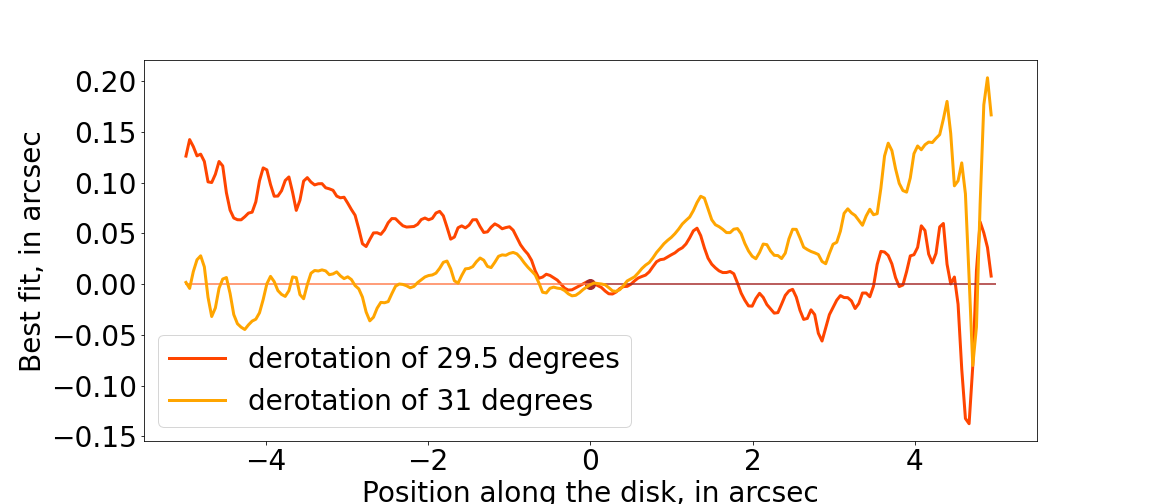}
\caption{Spine of the disk measured in the ClasImg image for the two position angles minimizing the disk slope in the northeast ($PA=31.0\deg$) and in the southwest ($PA=29.5\deg$). The
y-axis: departure from midplane in arcseconds. The x-axis: stellocentric distance in arcseconds. }
\label{fig:inclination}
\end{figure}

Measuring the position angle of the disk sets the base of its analysis. We used a similar method as in \citet{Lagrange2012}, \citet{Milli2014}, and \citet{Boccaletti2018} to extract the disk spine and derive the global position angle of the disk.
From the PCA image, we assumed a range of position angles around a guessed position ($\theta  = 30\deg$, $\Delta \theta  = 6\deg$, and $\delta \theta =0.1\deg$) and derotated the image by the complementary angle to position the disk near the horizontal orientation. The spine is defined as the departure from the mid-plane and is extracted by  perpendicularly fitting a Gaussian profile at each stellocentric distance on both sides of the star.
The local slope of the disk spine is measured at different positions along the spine, away  enough from the center  to avoid the bias from the self-subtraction induced by the PCA reduction. 
Averaging the values, typically ranging from 1.5$\arcsec$ to 4.0$\arcsec$, we searched for disk position angle that nulls down the slope.
In Figure \ref{fig:inclination}, we show the disk spine for the two values minimizing the slope in the northeasterly and southwesterly regions, which are
$31.0\deg$ and $29.5\deg$, respectively; these values are in relatively good agreement with the position angle measured at L band by \citet{Milli2014}. 
We adopted an averaged value of $30.0\deg\pm0.5\deg$, which appears discrepant with regard to the one used by \citet{Telesco2005}; however, applying the same method to measuring the PA in the VISIR data of 2004 and 2015 leads to a result of
$\sim33.3\deg\pm1\deg$ and $\sim34.4\deg\pm1\deg$, respectively.
As a reference, the most accurate value measured in scattered light owing to precise astrometric calibration is $29.2\pm0.2\deg$ \citep{Lagrange2012}.
The most notable difference of several degrees with the NEAR 2019 data could be related to the lack of an astrometric calibration procedure related to these data, especially since VISIR was moved to the VLT UT4 for the NEAR experiment. Still, this mismatch has no impact on the following analysis.

We 
note that in 
scattered light images, 
the midplane and the 
warp are observed as two 
distinct components owing to the edge-on 
orientation combined with the radiation pressure effect 
\citep{Golimowski2006, Lagrange2012}. 
\cite{Milli2014} has also referred to a warp in the L band, 
while, in fact, the disk image shows a single component that reveals a very similar trend to the one observed in the mid-IR: a single disk component with a misalignment on the two sides. 

Moreover, 
a larger PA in the mid-IR has also been reported by \citet{Pantin2005}, both in the N and Q bands. 

One possible interpretation can be drawn from temperature effects. 
Grains that are closer to the stars, thus, 
located inside the warp ($\lesssim$80\,au), 
dominate the global emission. In addition, the warp can be collisionally more active than the outer part of the disk due to the planet's gravitational influence, $\beta$ Pic b, onto the planetesimals, with a higher rate of small grains released, and with the latter acting as efficient emitters.
As a result, the mid-IR image of the disk would be essentially oriented along the warp.

\subsection{A clumpy structure}
\label{sec:disk_intensityclumps}

To analyze the clumpy structure of the disk, we converted the images to intensity units in Jy/arcsec$^2$, by taking into account the pixel area (a dilution factor of $2.10^{-3}$ corresponding to 1/pixel\_scale$^2$) and a photometric calibration which gives a total flux in the coronagraphic image of $1.1\times10^6$\,ADU for a point source of 1\,Jy. The surface brightness of the disk is obtained by integrating the intensity values for each pixel slice, of 12 pixels wide (0.54$''$), encompassing the disk thickness, and perpendicular to the mid-plane.

\begin{figure*}[t]
\centering
\includegraphics[width=0.8\textwidth]{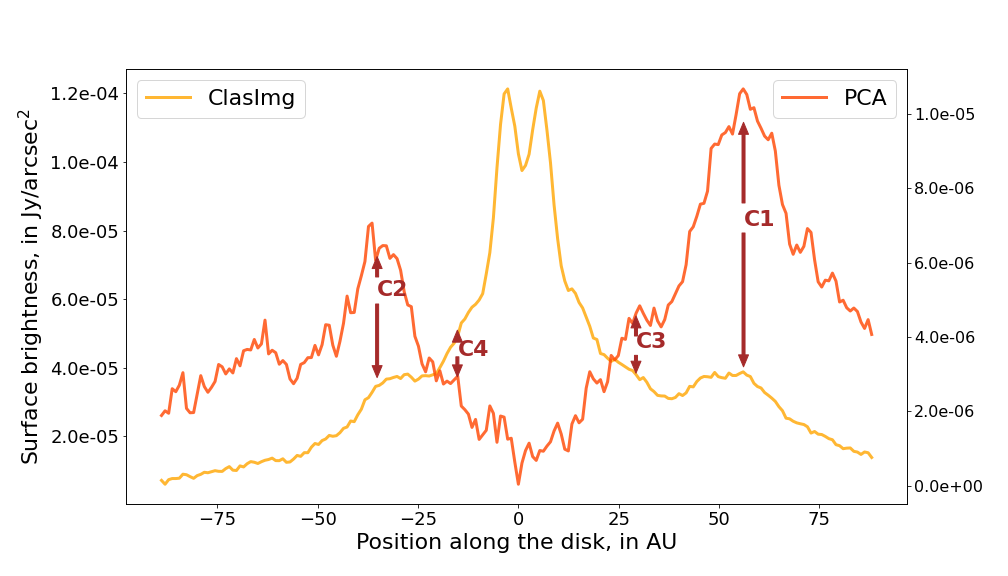}
\caption{Surface brightness profile of the PCA image (red), and ClasImg image (orange). The y-axes correspond to the  ClasImg on the left and the PCA on the right. Labels of the clumps have been added for convenience.}
\label{fig:noADI_PCA_intensity}
\end{figure*}

Figure \ref{fig:noADI_PCA_intensity} presents the resulting intensity profile along the disk for the ClasImg and PCA processing.
The clumps C1 and C2 can be identified in the surface brightness profiles, both in the ClasImg and PCA case. To a lesser extent, the hypothetical structures C3 and C4 are barely visible in the surface brightness.
As visible in Figure \ref{fig:PCAnoADI}, the PCA reduction  considerably attenuates the stellar contribution, impacting the photometry of the disk due to self-subtraction, with a stellocentric dependence.
We note that the clumps, in particular C1 and C2, are very elongated and that being seen edge-on there is an obvious degeneracy between radial and azimuthal extension (or possibly both). 
C1 has a projected width of $\sim$30\,au as seen in Figure \ref{fig:clumpintensities}, with a sharper width of $\sim$10\,au, corresponding roughly to the width of the peak in the PCA profile.

\section{Temporal evolution of the southwest clump C1}
\label{sec:disk_astrometryclump}

\begin{figure}[t]
\centering
\includegraphics[width=9cm]{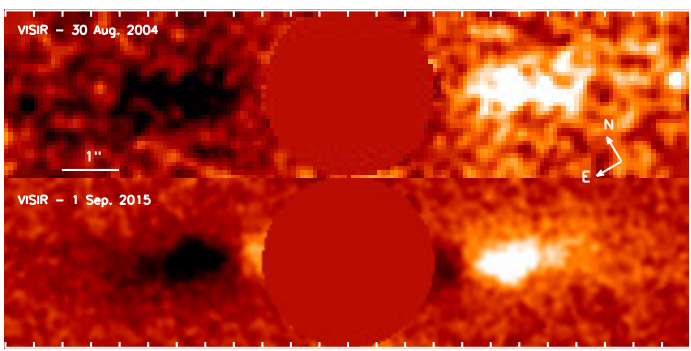}
\caption{Images of the centro-symmetrical subtraction of the disk at 11.25\,$\muup$m, obtained with VISIR in 2004 on VLT UT3 (top) and at 11.7\,$\muup$m obtained in 2015 (bottom). In both images, the disk has  rotated, respectively, 33$\deg$ and 34$\deg$ with respect to the north.}
\label{fig:disk2015}
\end{figure}

The $\beta$ Pictoris system has been observed in the mid-IR on several occasions with 8-m class telescopes since Dec. 2003, when C1 was first identified by \citet{Telesco2005}.  Given the 16-year baseline with respect to our NEAR observations, the study of its evolution becomes relevant and can potentially allow us to address questions around its origin. 
Table \ref{tab:datatable} presents the data available in the mid-IR that we used for comparison with our data set.  Both \citet{Telesco2005} and \citet{danli2012} located the position of C1 by performing a centro-symmetrical subtraction of the disk image, subtracting the emission in the fainter NE wing from the SW wing, and vice-versa. This yielded a residual emission, which was then assumed to be the main contribution of the clump. Indeed, in images without AO, the resolution is such that it is difficult to isolate the clump without this specific processing. For consistency reasons, we processed the 2004 and 2015 VISIR data similarly (Figure \ref{fig:disk2015}).  
In comparison, in the 2019 data, C1 is unambiguously detected and resolved, as seen in Figure \ref{fig:PCAnoADI}, without any particular processing. This  further emphasizes the efficiency of the use of AO along with a coronagraph.

\begin{figure}[t]
\centering
\includegraphics[width=0.5\textwidth]{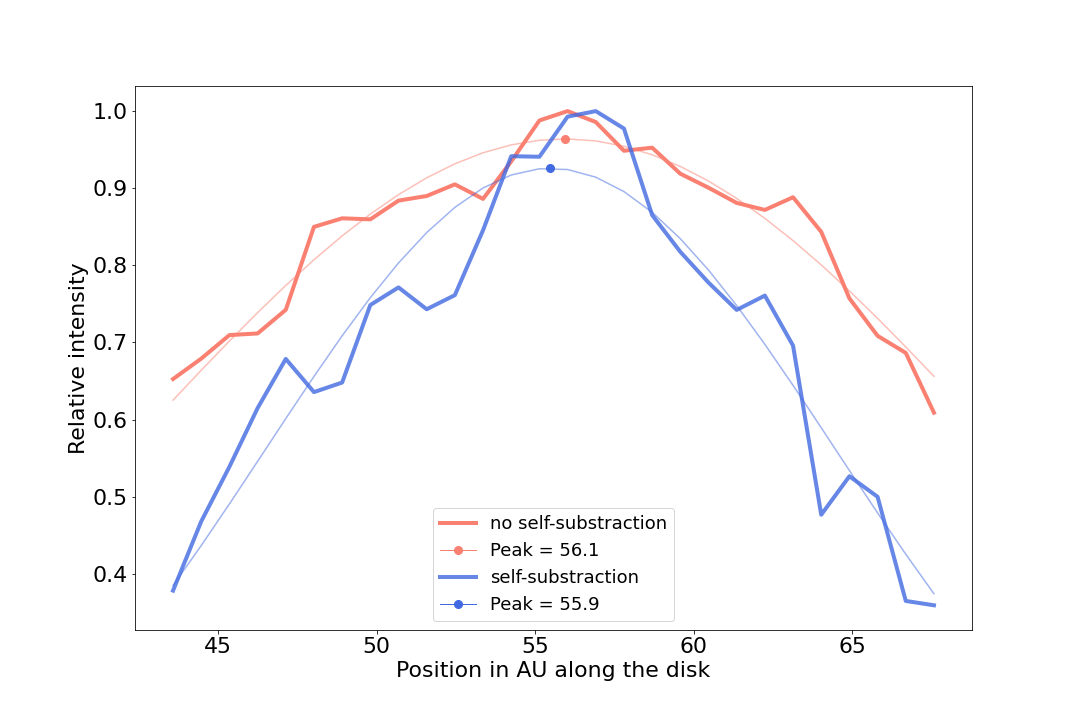}
\caption{Comparison of methods for measuring the position of C1 in the 2019 data. 
Red curve corresponds to the relative intensity from the PCA image, while the blue curve corresponds to the centro-symmetrical self-subtraction of the disk, from the same PCA image. Both methods provide the same result $\pm$ 0.2\,au. The position was measured by doing a Gaussian fit on the peak. The two curves have been normalized to make  the comparison more visible.}
\label{fig:ss_comp}
\end{figure}

\subsection{Impact of the centro-symmetrical subtraction}

Measuring the exact position of C1 from the intensity profile is not straightforward, given the width of the clump of several au. 
In the following, we assume that the position of C1 is driven by the intensity peak of this structure and we did not make any assumption on whether it is radially or azimuthally extended. Furthermore, as we used either PCA or centro-symmetrical subtraction, any stellar residuals or the main disk emission itself do not impact the determination of the C1 position.
In order to verify that the centro-symmetrical subtraction does not impact the position of C1, especially given the presence of the clump C2 and that the stellar contribution is substantial, we performed a measurement in the ClasImg 2019 image after applying the same centro-symmetrical subtraction. 

We found that the results of the two methods (centro-symmetrical subtraction in ClasImg vs. PCA ADI) are consistent within $\pm 0.1$\,au (corresponding to approximately $\pm0.1$ pixel), 
confirming the accuracy of the former method, as seen in Figure \ref{fig:ss_comp}. Hence, we did not apply the centro-symmetrical subtraction to determine the location of C1 in the 2019 data. \\

\begin{figure}[t]
\centering
\includegraphics[width=0.45\textwidth]{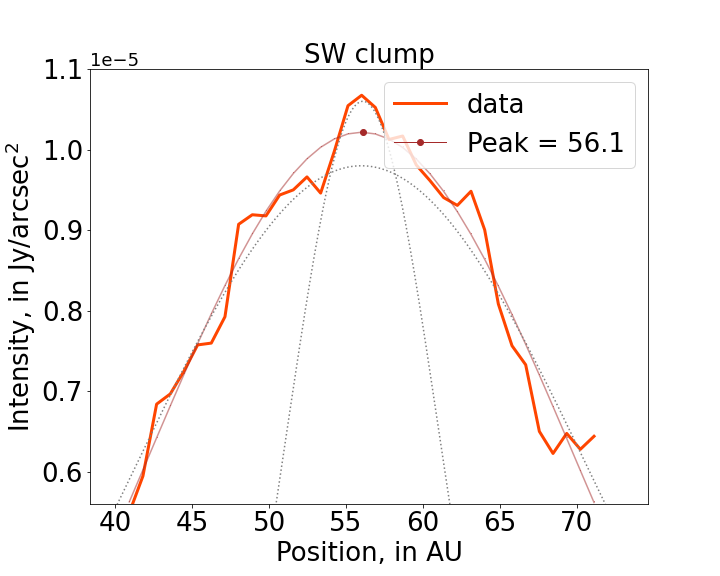}
\caption{Position of C1 in the PCA image of the 2019 data. Two Gaussian curves have been added in dashed lines to show the substructures within the disk and the brown line corresponds to the final fit.}
\label{fig:SWgauss}
\end{figure}

\subsection{Position and error bar estimations}
\label{sec:errorbars}
We considered the following sources of uncertainty in measuring the position of C1 and we describe the resulting errors for each of these contributions to the 2019 data. \\

The first source of error comes from estimating the \textbf{position of the star in the image}. We performed this estimation by doing a Gaussian fit of the center of the image in the ClasImg process, compared to the center of the image. The residual error is coming from the derotation of the images in the ClasImg process, when assuming the star is centered on the image. We estimated a resulting 0.2-pixel error corresponding to the cumulative error in the ClasImg sequence. This includes QACITS pointing control error, which centers the star on the coronagraph within $\sim$ 0.02\,$\lambda/D$.\\

The second error comes from the measurement of the \textbf{orientation of the disk}: repeating the process of measuring the clump position for the range of PA ($30\deg\pm0.5\deg$), we found a dispersion lower than 0.3 pixel. 

The third source of error lies in the \textbf{Gaussian fitting of the clump}: to accurately measure the position of the clump, we limited the range of distance from the star where we fit a Gaussian profile (from around $2''$ to $4''$) centered approximately on C1, as seen in Figure \ref{fig:SWgauss}).
The resulting error is 0.1 pixel. \\

Finally, the \textbf{data reduction method} induces biases. First, the PCA reduction introduces a bias leading to a photometric error; furthermore, the centro-symmetrical subtraction performed for the 2004 and 2015 data adds up. To estimate the bias introduced by the PCA reduction, we compared the position of C1 with the one found with the ClasImg reduction by performing a centro-symmetrical subtraction, resulting in a shift of 0.1 pixel.\\

Taking these factors into account, the total uncertainty for the 2019 data corresponds to the quadratic sum of each contribution, which equals 0.34 pixels (equivalent to 0.3\,au). 

With regard to the 2015 data, the same sources of uncertainty were considered for the error estimation, except that QACITS was not used for the centering. The centering error is of $\sim$0.3 pixels, due to the bright PSF making it more challenging to accurately locate the star center as compared to the 2019 data. As for the position angle of the disk, the error is estimated to be lower than 0.3 pixels as well, as long as the same method as the 2019 data was applied. The error induced by the starting points of the Gaussian fit is 0.1 pixel. Here, the centro-symmetrical subtraction is another contribution to the error, which is estimated to be 0.1 pixel (cf. Figure \ref{fig:ss_comp}). The resulting uncertainty corresponds to 0.45 pixel, corresponding to 0.4\,au.

The 2004 data follows the same error estimation calculations as the 2015 data, with a difference coming from the pixel scale of the detector being different (0.075 arcsec/pix, instead of 0.045 arcsec/pix for 2015 and 2019, respectively, following a sensor upgrade). The resulting uncertainty corresponds to 0.6\,au. 

In conclusion, the C1 projected distances for the 2004, 2015, and 2019 data are $52.7\pm0.6$\,au, $55.5\pm0.4$\,au, and $56.1\pm0.3$\,au, respectively. These results are summarised in Table \ref{tab:datatable}.

\subsection{Temporal evolution of C1's projected distance}
\label{sec:C1positions}
\begin{figure*}[t]
\centering
\includegraphics[width=0.8\textwidth]{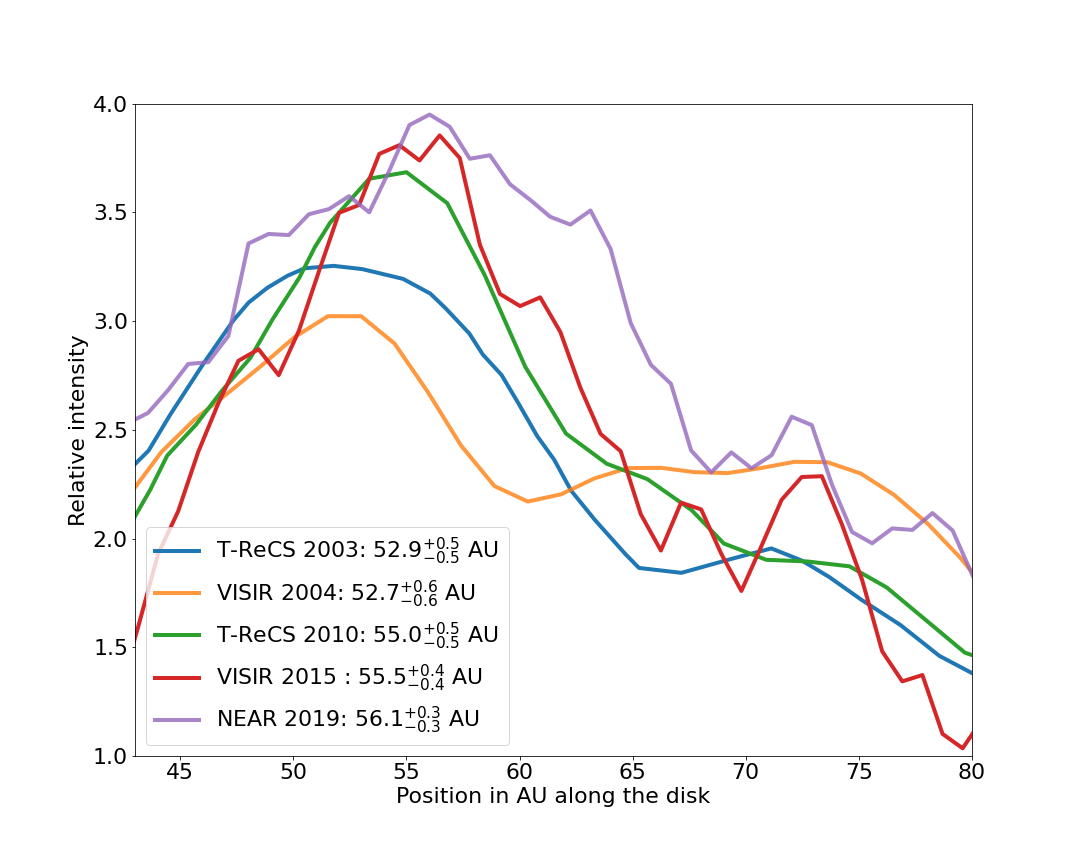}
\caption{Evolution of the projected distance and relative intensity profile of C1 over the years, from different instruments (colour-coded in the legend). The intensity for the VISIR and NEAR data has been set to scale with the T-ReCS data by \cite{danli2012}.}
\label{fig:clumpintensities}
\end{figure*}

Figure \ref{fig:clumpintensities} presents the intensity profiles of the disk for each epoch around the location of C1. The
T-ReCS measurements from 2003 and 2010 were obtained from the plots in \citet{Telesco2005} and \citet{danli2012}, after updating the star's distance to 19.63.

The apparent projected separation of the clump is clearly increasing over time, suggestive of a global outward motion of C1. 
This is even more obvious when plotting its measured position versus time on a 16-year baseline, as in Figure \ref{fig:clumporbit} (red circles for 2003 and 2010 data, red squares for 2004, 2015, and 2019 data). 
This outward motion of the projected distance is likely to be slowing down over time, indicating that the clump should be coming close to its maximum elongation.

We note that the overall profile of C1 has a two-mode shape with a nearly Gaussian part inward and a plateau outward, indicative that the actual three-dimensional shape of the clump is more complex. In particular, the clump can be azimuthally and/or radially extended. In the latter, the differential Keplerian rotation will modify the projected intensity profile with time, an effect which could already be suspected in Figure \ref{fig:clumpintensities}.
Furthermore, we considered studying the evolution of the size of the clump, however, the current data do not allow us to accurately quantify such evolution.
Follow-up observations with similar or better angular resolution than NEAR will be key to future investigations of the actual morphology of the clump.
 
Carrying out a quantitative comparison of  the disk intensity profiles is impractical, given the absolute flux is not known accurately enough in each dataset. 
For that reason, we normalized the profiles in Figure \ref{fig:clumpintensities} to roughly match the SW side intensities beyond 65\,au, for the purposes of visualization. 


\subsection{C1's orbital radius}
\label{sec:disk_orbitclump}

To set some constraints on the orbit of C1 -- given the scarcity of data points, the small fraction of the orbit coverage, and the rather large error bars with regard to the clump location -- we assumed a simplified configuration in which the clump orbit is circular and perfectly edge-on. 
Therefore, we purposely excluded the possibility of elliptical orbits since they would bring on too many solutions.
In that case, the Keplerian angular velocity is expressed as:
\begin{equation}
\Omega_K=\sqrt{\frac{GM_{\star}}{a^3}}
,\end{equation}

with $a$ being the semi-major axis and $G$ being 
the gravitational constant. The projected separation, $x_0$, at the first epoch is given by: 

\begin{equation}
    \cos(\theta_0)=\frac{x_0}{a}
,\end{equation}
with $\theta_0$ defining the 
angle between the clump and the direction perpendicular to the line of sight.
Since the Keplerian speed is constant for a circular orbit ($\Omega_K=\Delta \theta/\Delta t$), the projected separation, $x[t,a]$, as a function of time for a given orbital radius is expressed as:

\begin{equation}
    x[t,a]= a \cos\left(\theta_0 - t.\sqrt{\frac{GM_*}{a^3}} \,\right)
.\end{equation}

\begin{figure}[t]
\centering
\includegraphics[
width=9cm, trim=2cm 1cm 2.5cm 0cm, clip]{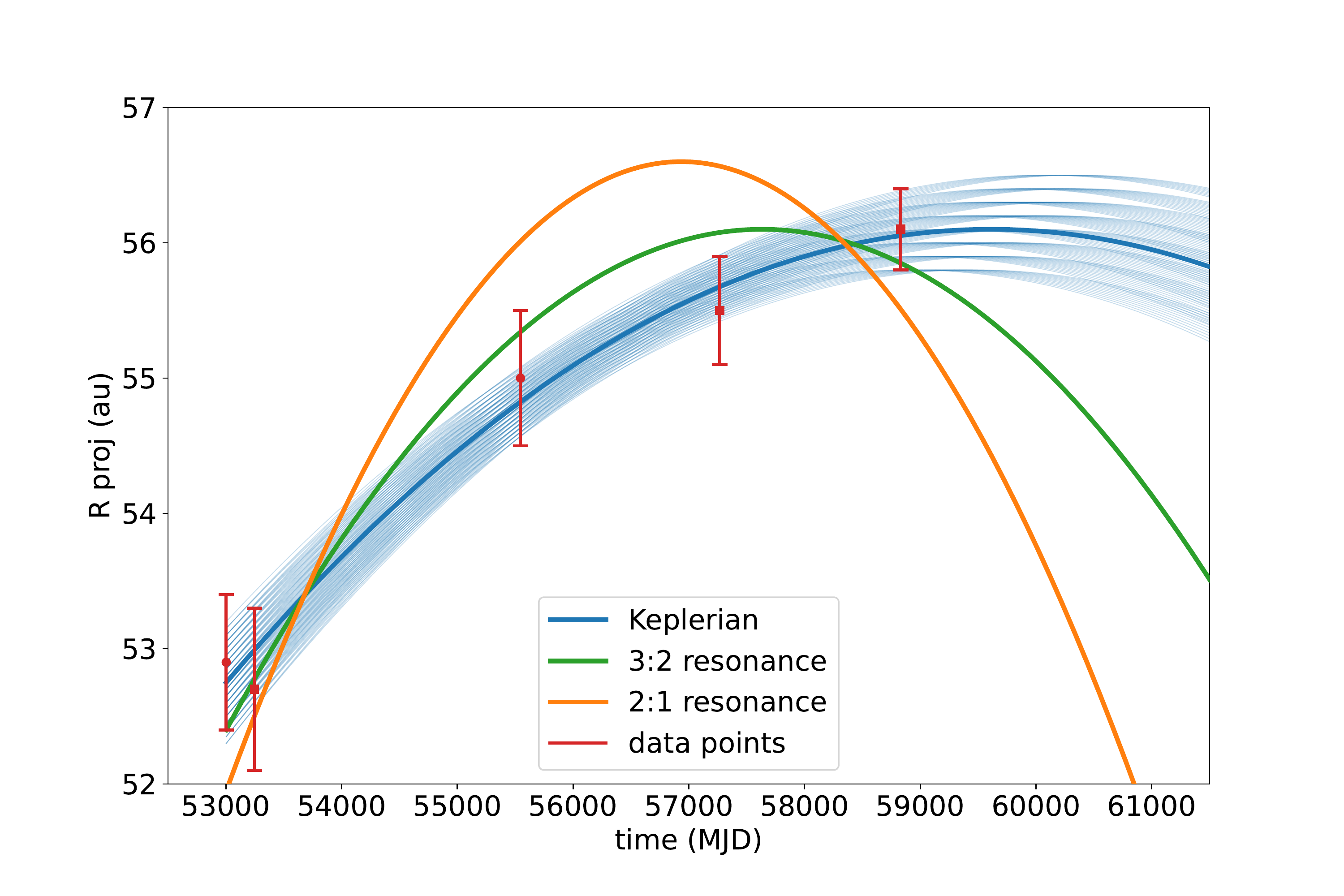}
\caption{Evolution of the position of C1 over the 16 years of observations (red squares for VISIR and NEAR, red circles for T-ReCS). The best Keplerian model is overlaid in blue together with the 1-sigma dispersion (light blue), corresponding to $R=56.1^{0.4}_{-0.3}$\,au. Also displayed: best models fitting the data for a 2:1 (orange) and 3:2 (green) resonances.}
\label{fig:clumporbit}
\end{figure}

To estimate the best models matching the data given the error bars, we generated a grid of 2000 models of two parameters, $a$ and $x_0$, with the following priors: $55-60$\,au (step 0.1\,au) and $51.9-53.9$\,au (step 0.05\,au), respectively. 
As a result of a $\chi^2$ minimization, we were able to constrain the semi-major axis of C1 
to $a=56.1^{+0.4}_{-0.3}$\,au (and $x_0=52.8^{+0.5}_{-0.5}$\,au). More details are given in Figure \ref{fig:clumporbit}).

\citet{Telesco2005} measured a position for C1 of 52\,au in Dec. 2003, while \citet{danli2012} obtained 54\,au, then speculated a displacement of roughly $2.0^{+0.6}_{-0.6}$\,au. The latter authors concluded that the clump is moving at Keplerian velocity, corresponding to an orbital radius of $54.3^{+2.0}_{-1.2}$\,au. When considering a stellar distance of 19.63\,pc \citep{Gaia2021} instead of 19.28\,pc, the orbital radius changes to $54.7^{+2.0}_{-1.2}$\,au, according to Figure 8 in \citet{danli2012}. This is a value that ought to be compared to the obtained values  described in the previous section.
Therefore, our measurement
is in agreement, within the error bars, with this revised value. 
We note that this value corresponds almost exactly to the projected separation of the clump in the 2019 data, hence, the clump is supposed to be at its maximal elongation.  


\section{Other clumps and comparison with ALMA}
\label{sec:disk_otherclumps}

\begin{figure}[ht]
\centering
\includegraphics[width=0.45\textwidth]{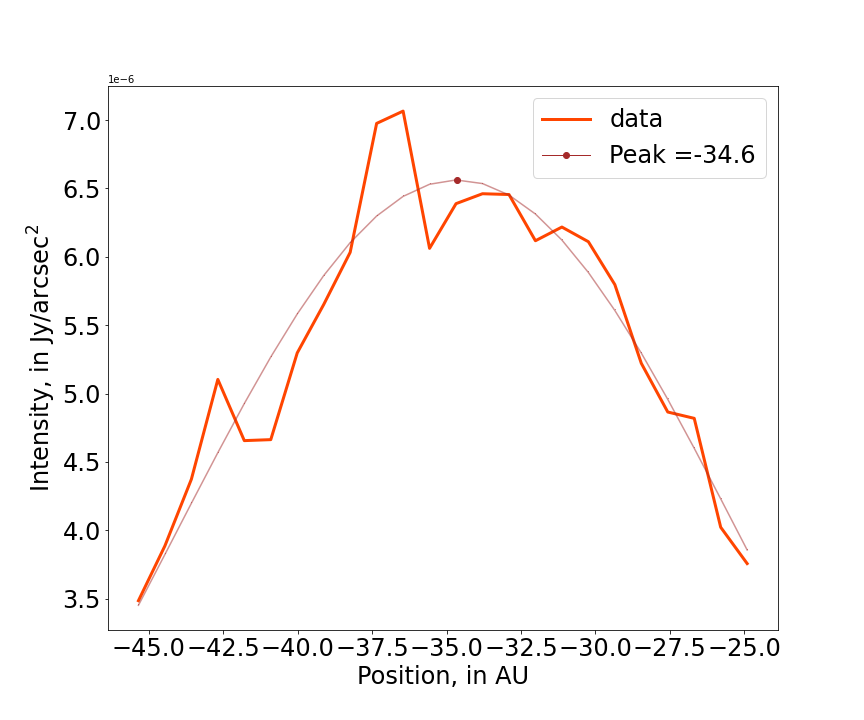}
\caption{Relative intensity along the disk in the PCA image, focused on C2. The intensity values were kept the same as in Figure \ref{fig:clumpintensities}.}
\label{fig:NEclump}
\end{figure}

As mentioned in Sect. \ref{sec:disk}, the 2019 NEAR images of the disk feature additional clump-like structures. We measured the projected separations 
of these structures following the approach detailed in Sect. \ref{sec:disk_astrometryclump} for the clumps C2, C3, and C4. 
We obtained, respectively,  $-35.2^{+0.3}_{-0.3}$\,au, $29.3^{+0.3}_{-0.3}$\,au, $-15.2^{+0.3}_{-0.3}$\,au for the clumps C2, C3, and C4 positive towards SW and negative towards NE. 
Table \ref{tab:clumps} summarizes the clumps positions. 
The intensity profile of C2 is shown in Figure \ref{fig:NEclump}.
Similarly to the C1 clump, we fit a Gaussian curve on C2, to estimate its position making the assumption that the profile of the clump is Gaussian yielding
a width of $\sim$15\,au.

\begin{table*}[t]
\centering
\begin{tabular}{lllll}
\hline

                       \textbf{Clumps}   & \textbf{C1}   & \textbf{C2}    & \textbf{C3}   & \textbf{C4}   \\ \hline
                          \hline
\textbf{Projected separation (au)} & 56.1$\pm$0.3 & -35.2$\pm$0.3 & 29.3$\pm$0.3 & -15.2$\pm$0.3 \\ \hline
\end{tabular}
\caption{Projected separations of the clumps observed with NEAR. Negative signs are for the NE side of the disk.}
\label{tab:clumps}
\end{table*}


\cite{Dent2014} and \cite{Matra2017} observed the $\beta$ Pictoris disk at submillimeter wavelengths with the Atacama Large Millimeter/submillimeter Array (ALMA), presenting the spatial distribution of the CO gas for the transitions J=3-2 (resolution of 15\,au) and J=2-1 (resolution of 5.5\,au). Both studies identified two clumps respectively to the SW ($\sim 50$\,au) and the NE ($\sim 30$\,au).

\begin{figure}[t]
\includegraphics[width=0.54\textwidth]{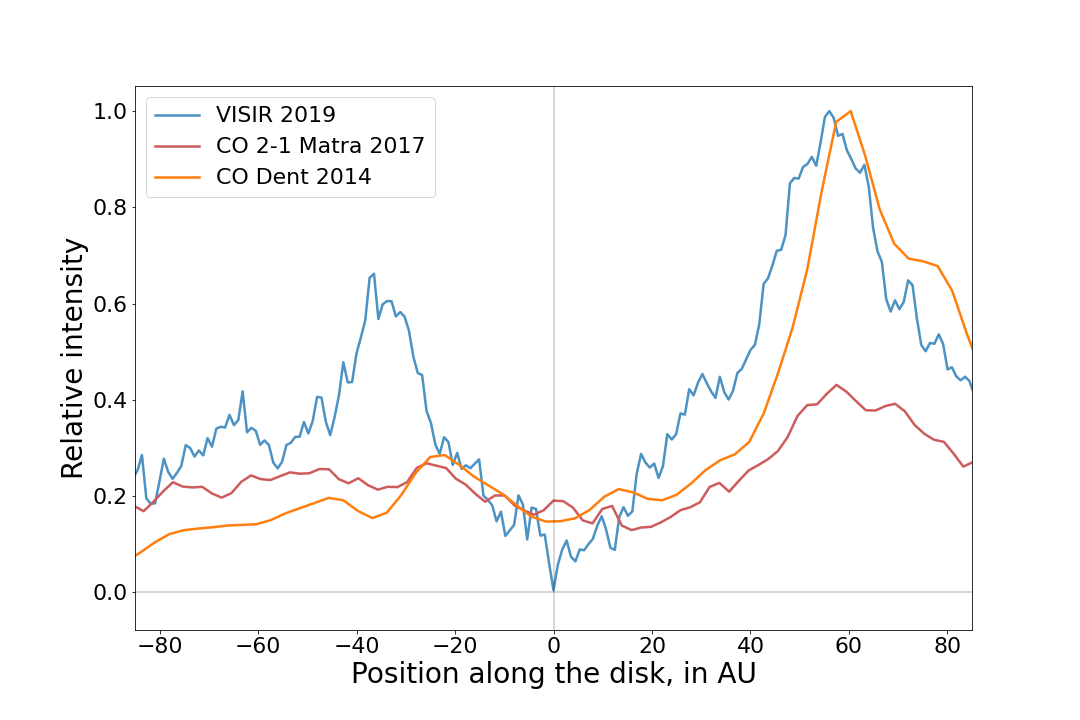}
\caption{Projected distribution of CO lines flux obtained from \cite{Dent2014} and \cite{Matra2017}, superimposed with the radial flux distribution of the disk at 12$\muup m$ with VISIR. The VISIR and CO 3-2 \citep{Dent2014} curves have been normalized by their maximum.}
\label{fig:CO}
\end{figure}

{To assess to what extent these CO clumps match those we have identified here, we show (in Figure \ref{fig:CO})} the superimposition of the {projected} intensity profiles 
for the NEAR 2019 data, together with the ALMA data. For consistency, we used the star's distance of 19.63\,pc for these three observations, thus including potentially slight differences with the two aforementioned papers regarding the positions of the structures. 
Although we detect with VISIR a well-resolved clump (C2) in the NE part of the disk, the correspondence with the CO gas distribution is not very conclusive. Not only does the clump on the NE side in ALMA images peak at about 25\,au, as opposed to 33\,au in the VISIR image, but it is also related to a much broader projected structure, especially in CO J=3-2 ($\sim$70\,au), than at 11.25\,$\muup$m. On the SW side, on the contrary, there is a good match between the projected location of C1 and that of the CO gas clump.
A small offset ($\sim$6\,au) of C1 between ALMA and VISIR observations  is visible in Fig. \ref{fig:CO}, but this cannot be attributed to the clump orbital motion according to the analysis in Sect. \ref{sec:disk_orbitclump}, possibly implying a slightly different distribution of the dust and gas components.
However, we note that \cite{Matra2017} argued that while the projected location of the CO clump is around 50\,au, its deprojected stellocentric distance (deduced from position-velocity (PV)  diagrams of CO intensity)
peaks, in reality, at $\sim85\,$au. This seems to contradict the results of our orbital analysis (Sec.~\ref{sec:disk_orbitclump}), which  the semi-major axis of the dust clump is constrained to lie around 56\,au. At face value, this would suggest that we are witnessing two different clumps whose projected positions happen to coincide. Such a conclusion should, however, be taken with great caution. The PV diagrams of the CO lines do indeed also suggest that the gas clump should be radially extended, spanning $\sim$100\,au for the orbital radius. If the dust clump was to have a similar radial extension, then it could appear brighter (because of higher temperatures) at its inner edge, while the peak CO luminosity could be located further out, thus explaining the apparent discrepancy. In addition, the different transitions of CO gas will be more or less excited depending on temperature and the density of collisional partners \citep[in non-LTE as is the case for $\beta$ Pic,][]{Matra2017} and they are not necessarily representative of the underlying CO or dominant gas species spatial distributions \citep[e.g.,][]{Kral2016}. These questions clearly go beyond the scope of the present paper, and we leave this to be an open issue for investigation in future studies.

\section{Origin and fate of C1}
\label{sec:discussion}

As shown in Sect. \ref{sec:disk_orbitclump}, the location of C1
over the 2003-2019 period is compatible with a circular Keplerian orbit 
with a $56.1^{+0.4}_{-0.3}$\,au semi-major axis.
{ During the review process for the present paper, \cite{Han2023} published a study arguing that the main SW clump (C1) ought to be stationary, which seems to contradict our conclusion. We note, however, that, while \cite{Han2023} constrained the maximum displacement of the clump to, indeed, be less than 0.2\,au over a 12-year period at the $1\sigma$ level, this value increases to 11\,au at the $3\sigma$ level, which would be compatible with our own results (see Sect.~\ref{sec:disk_orbitclump}). We also note that differences could arise from  different approaches for pinpointing the clump's location: we constrained it by looking for the peak luminosity location while \cite{Han2023} constrained it by performing a fit of the whole projected profile of the clump. Lastly, the present study considers a longer time baseline, with the 2019 NEAR data extending it to 16 years instead of 12.
Keeping in mind these possible caveats, we go on to review some possible scenarios for explaining the clump's motion over time below.}

\subsection{Giant impact}
Since its detection by \cite{Telesco2005}, several explanations have been proposed for the presence of C1. The first is the catastrophic disruption of a large ($\sim100\,$km) planetesimal \citep{Telesco2005,danli2012}. As demonstrated by \cite{Jackson2014} and \cite{Kral2015}, the fact that the produced collisional debris are placed on eccentric orbits, all passing through the location of the initial break-up, produces a long-lived bright clump at this location. However, this clump stays at a fixed position with respect to the star, which does not agree with the observed motion of the clump over a 16-year interval. 
The only way a catastrophic disruption leads to a moving clump is if it is observed in the immediate aftermath of the break-up, before the debris had time to perform a complete orbit \citep[see, e.g., Fig. 7 of][]{Jackson2014}. While this possibility cannot be ruled out, it is  very unlikely given the fact that such large disruptive events  are known to be relatively rare \citep{Wyatt2016}.

\subsection{Collisional avalanche}
Alternatively, \cite{danli2012} considered the possibility that given the large radial extension of the clump and the fact that it might contain sub-micron grains, it is the signature of a so-called collisional avalanche. 
This is a collisional chain reaction triggered by outward moving unbound small grains produced by the break-up of planetesimals closer to the star \citep{grigorieva2007}. However, the duration of an avalanche event is relatively short, on the order of $\sim0.3\,t_{orb}$, again requiring the assumption that we are witnessing the immediate aftermath of a large planetesimal break-up \citep{thebault2018}. However, contrary to the "local" giant disruption scenario considered before, the avalanche-triggering planetesimal break-up would occur much closer to the star (at a typical asteroid-belt location), in regions where such events could be less rare, and would require the breaking up of a smaller body \citep{thebault2018}.

\subsection{Resonance trapping by a planet}
Analyzing the characteristics of the CO clump discovered at roughly the same projected location as the dust clump, \cite{Dent2014} and \cite{Matra2017} also ruled out a giant disruption event and favored instead a scenario in which the clump is the result of resonance trapping of CO-producing planetesimals by a planet moving with a Keplerian orbit.
Such a scenario would imply that the clump moves at the angular velocity of the planet, that is, significantly faster than the expected Keplerian speed at the location of the clump. We tested this hypothesis by fitting the 2003-2019 clump positions when assuming that it moves at the angular speed of a planet with which it is in either a 3:2 or 2:1 resonance \citep[the two cases considered by][]{Matra2017}. 
As shown in Figure \ref{fig:clumporbit}, a 2:1 resonance can be confidently ruled out (reduced minimal $\chi_\nu^2=6.3$), while a 3:2 case might be marginally possible (reduced minimal $\chi_\nu^2=1.4$) given the error bars.
Still, the scenarios with resonances are significantly worse than when assuming the local Keplerian orbit (reduced minimal $\chi_\nu^2=0.22$, Figure \ref{fig:clumporbit}). We note that for this 3:2 resonant scenario, any new observation of the clump location should easily settle the validity of this hypothesis.

\subsection{Planet's Hill sphere or trojans}
The most likely hypothesis we are left with is thus that of a dust clump that orbits at the expected local Keplerian speed and should be relatively long-lived in order to be observed. 
If this clump is linked to the presence of a yet-undetected planet, then it could be either circumplanetary material within the planet's Hill radius or Roche lobe or, alternatively, material trapped in the corotating L4 or L5 Lagrangian points, in a Trojan-like configuration. \cite{Telesco2005} estimated the total mass of dust in the clump to be around $4\times10^{20}$\,g, which is much less than the estimated mass of the Jupiter Trojans, $\sim6\times10^{23}$\,g \citep{Jewitt2000}. 
However, $4\times10^{20}$g is the estimated mass of dust (typically $\leq1\,$mm), whereas the estimated mass of Trojans is that of $\geq1\,$km objects. If we assume that the observed dust is produced by a collisional cascade starting at $\sim1\,$km object, with a differential size distribution following a power law of index $q=-3.5$, then we get an extrapolated total mass of $\sim4\times10^{23}$\,g, which is roughly comparable to the mass of km-sized Jupiter Trojans.{ There is, however, a potential issue with this scenario if the dust clump has the same radial extent as that of the clump seen in CO.}
The relative radial width of the stable region around the L4-L5 points should indeed not exceed $\sim10$\% \citep{liberato2020}, whereas it is at least $\sim50$\% for the observed CO clump \citep{Matra2017}.{ As discussed in Sec.~\ref{sec:disk_otherclumps}, the correspondence between the dust and CO clumps is a complex issue that is left to future investigations. We remain careful as we stress that the Trojan scenario might be challenged by a radially broad dust clump. } 

{ Another possibility is that the clump is confined within the Hill sphere surrounding a planet. In this case, the minimum mass of the putative planet can be derived from the size of the clump, using Equation \ref{eq:hillequation} and assuming that the clump is smaller than $R_{Hill}$. For a typical clump size of $\sim$10\,au, the mass of the planet having a Hill sphere of this size is of the order of 3$M_J$. The problem is that, at a distance of $\sim50\,$au, such a massive planet would have had a 99\%  probability of having been detected by \cite{lagrange_unveiling_2020} with a combination of radial velocity and imaging. 
This problem could be overcome if the clump is optically thick ($\tau>1$) and, thus, it would be hiding the planet's photosphere. For a 10\,au wide clump made of the smallest possible grains ($\sim2\mu$m), the $\tau>1$ criteria  would lead to a minimum clump mass of $\sim5\times10^{25}$g, which corresponds to the mass of a Moon-sized object.}

\subsection{A vortex} 

In the hypothetical case that the gas and dust clumps are co-located, a scenario that could allow for a somewhat large radial extent \citep[as deduced from observations of the gas velocity][]{Matra2017}, while also explaining that the gas orbits at the local Keplerian velocity are those characterizing a vortex. Vortices have been extensively studied, both analytically and numerically, in the context of younger protoplanetary disks, yet they have never been proposed as an explanation for clumps in debris disks. There are several ways to generate vortices in disks, but large-scale vortices are most often thought to arise from the Rossby-wave instability (RWI). This instability can set in when there is a radial minimum in the gas potential vorticity, which, in practice, can happen when there is a radial maximum in the gas pressure (see, e.g., the review by \cite{LovelaceRomanova2013}).

Radial pressure maxima could well occur in debris disks. The pressure maximum could be related to: 1) a Saturn-like planet or more massive creating a gap with a natural pressure maximum at its outer edge \cite[just like in a gas-rich protoplanetary disk; see, e.g.,][]{Hammer2021}; 2) the presence of a clump of solids (similar to that observed in $\beta$ Pic) that releases gas and naturally creates a pressure maximum. Large-scale vortices can also survive for thousands of orbits, namely, millions of years at $>50$\,au \citep{Hammer2021}, making it possible to observe in a $\sim$20 Myr-old system. We note, however, that the typical lifetime of vortices depends, among other things, on the gas turbulent viscosity, with smaller turbulent viscosities favoring longer-lived vortices \cite[see, e.g., the discussion in Sect. 4.2 in][]{Baruteau2019}. The amount of turbulence in debris disks is not yet observationally constrained but it could be very high in low-gas mass ionized systems and lower in more massive disks \citep{Kral&Latter16}.

According to current RWI models, the smallest dust with a Stokes number of $\sim$1 ($\sim$ $\mu$m dust) is expected to have a tendency to concentrate near the center of the vortex, but the effect of radiation pressure, which cannot be neglected in debris disks, has not been taken into account in RWI models and may alter this conclusion. We also note that when the dust-to-gas mass ratio in the vortex becomes greater than 0.3-0.5, the vortex may be destroyed \citep{Crnkovic-Rubsamen2015}. In the case of $\beta$ Pic, \cite{Kral2016} calculated that the dust-to-gas mass ratio is greater than 1 beyond 20\,au, so this effect of dust feedback on the gas vortex may be relevant here.

The presence of a vortex would also solve another as-yet unexplained phenomenon, namely, that the neutral carbon gas observed with ALMA is not axisymmetric (as predicted by the models) but clumpy, similarly to what is observed for CO \citep{Cataldi2018}. Indeed, according to current models, CO should photodissociate in less than one orbit, which may explain why it is clumpy \citep{Matra2017}, but the carbon that is created due to the photodissociation of CO should instead become axisymmetric rapidly on a time scale of a few orbits. On the contrary, if the gas forms a vortex, both the carbon and CO are indeed expected to be clumpy. The large width of the clump is also in line with the idea of a vortex in the gas. An RWI-induced vortex has indeed a radial width that is typically twice the local pressure scale height \citep{Baruteau2019} and can thus reach tens of au, as observed for $\beta$ Pic.

We also note that a significant brightness asymmetry is observed in the near IR  \cite[e.g.][]{Apai2015} and a clump in the mid-IR \cite[][and this study]{Telesco2005} but not in the mm \citep{Matra2019}. This could also be explained by the presence of a vortex trapping only the smallest $\mu$m-sized grains (with a Stokes number $\lesssim$ 1), while the largest mm-grains would remain unperturbed by the vortex and retain a near axisymmetric spatial distribution. These statements should be further analyzed via numerical simulations in a dedicated study.

Although further numerical simulations for the specific case of debris disks are needed to strengthen our conclusions, vortices are compelling contenders for explaining the results of observations and they deserve more attention in the context of $\beta$ Pic, as this would allow us to explain (for the first time) the observations of CO and carbon gases as well as the observations of dust in the near- and mid-IR, as well as in the mm. For all these reasons, we suggest that it is {a viable scenario that needs further theoretical and observational testing}.

If we expect C1 to be caused by a vortex formed at the outer edge of the annular gap of a planet, we could offer an idea of where the planet would be located according to its mass. A vortex formed at the outer edge of a planet's gap is typically located within a few (5-10) Hill radii of the planet. For a Jupiter-mass planet around a Sun-mass star, the vortex will have a semi-major axis $\sim1.5\,a$ (see for instance Figure 2 by \citealt{Baruteau2019}). In our case, the hypothetical planet would be located around 37\,au {(assuming a dust clump centered at $\sim$50 au)}, which is between C3 and C1.

\section{Conclusion}
\label{sec:conclusion}

This paper presents the first high contrast imaging data in the mid-IR of the $\beta$ Pictoris system, observed with NEAR. 
Here, we summarize the main results of our analysis of these data, along with a comparison with previous observations. 

\begin{itemize}
    \item The planet $\beta$ Pictoris b was not detected with NEAR. However, by taking into account the transmission of the coronagraph, we derived the planetary flux upper limit from the contrast curve at 5$\sigma$. We collected
    spectro-photometry from different instruments and presented a combined spectrum the first spectra %
    of $\beta$ Pictoris b acquired with SPHERE during several epochs. The upper limit of the NEAR data does not allow us to put meaningful constraints on atmospheric scenarios.
    \item Given the upper limit on the planetary flux, we investigated which corresponding amount of dust, present around the planet, could reproduce such a limit.
    Although this quantity scales with the distance to the planet, we concluded that the presence of a dust cloud around the planet was unlikely. If dust particles were located at the Roche radius, 
    it would correspond to the collisional debris of a 5\,km size asteroid.
    
    \item The disk is uniquely resolved in these mid-IR data, allowing us to identify structures that were never observed prior to those observations. The southwesterern clump, previously reported in the literature, is distinctively detected in the NEAR data set, at $56.1^{+0.3}_{-0.3}$\,au. On the northeastern side of the disk, there is a clear detection of a new clump at $-35.2^{+0.3}_{-0.3}$\,au. We note the possible presence of two other clumps at $29.3^{+0.3}_{-0.3}$\,au at $-15.2^{+0.3}_{-0.3}$\,au. Further observations will be required to confirm their existence. 
    
    \item The southwest clump was observed several times since its discovery in 2003, with T-ReCS and VISIR. The 16-year baseline of observations, with five observing sequences, allowed us to assess the motion of the clump over time, and to confirm a Keplerian behavior. 
    {This result is based on the assumption that the clump has a circular orbit, which would be challenged in the case of an elliptical orbit.}
    \item We investigated qualitatively different origins for the southwest clump, in particular, the possibility that it could be the Hill sphere of a yet-to-be detected planet with a maximum mass of 3M$_J$.
    %
    Given the fact that it is in motion, we ruled out the scenario that this clump is the result of a giant impact. We have provided arguments for a scenario where the clump would be a dust-trapping gas vortex, based in particular on { the possible superimposition of} the southwest clump and the CO clump seen with ALMA.
    
\end{itemize}

The $\beta$ Pictoris system has been extensively studied at different wavelengths and it serves an archetypal system for our understanding of planet-disk interactions and planetary formation. 
This study shows that the use of adaptive optics, along with a coronagraph in the N band, does bring a considerable improvement with regard to the data quality.
The NEAR data have allowed us to put constraints, for the first time, on the presence of circumplanetary material around a directly imaged planet. We could indeed expect giant planets to have circumplanetary disks, as all the giant planets in our solar system   do indeed have some. Similarly, we detected disk structures that had never been observed before.
Further observations are needed to confirm some of those structures and understand their origin. Likewise, further observations of C1 are needed to track its evolution and confirm its origin, and to put better constraints on the evolution of the size of the clump. 
The {\it James Webb Space Telescope} will provide unique mid-IR data that will significantly help us in improving our understanding of the famous $\beta$ Pictoris system.

\begin{acknowledgements}
We want to thank the ESO, the Breakthrough Foundation, and everyone involved in the NEAR project. The observations were carried out under the ESO program id: 60.A-9107(K).
We want to thank the referee for the constructive feedback that contributed to improving the quality of this paper. 
N.S. acknowledges support from the PSL IRIS-OCAV project.
This project has received funding from the European Research Council (ERC) under the European Union's Horizon 2020 research and innovation programme (COBREX; grant agreement \#885593).
French co-authors also acknowledge financial support from the Programme National de Plan\'{e}tologie (PNP).
We thanks J. Chilcote for sharing the GPI spectrum of $\beta$ Pic b.
C.D. acknowledges financial support from the State Agency for Research of the Spanish MCIU through the ''Center of Excellence Severo Ochoa'' award to the Instituto de Astrof\'isica de Andaluc\'ia (SEV-2017-0709) and the Group project Ref. 
N.H. was partially funded by Spanish MCIN/AEI/10.13039/501100011033 grant PID2019-107061GB-C61 and No. MDM-2017-0737 Unidad de Excelencia {\em Mar\'{\i}a de Maeztu} - Centro de Astrobiolog\'{\i}a (CSIC-INTA).
PID2019-110689RB-I00/AEI/10.13039/501100011033.

\end{acknowledgements}

\bibliographystyle{aa}
\bibliography{main}

\begin{appendix}

\section{Transmission of the coronagraph}
\begin{figure}[t]
\centering
\includegraphics[width=0.45\textwidth]{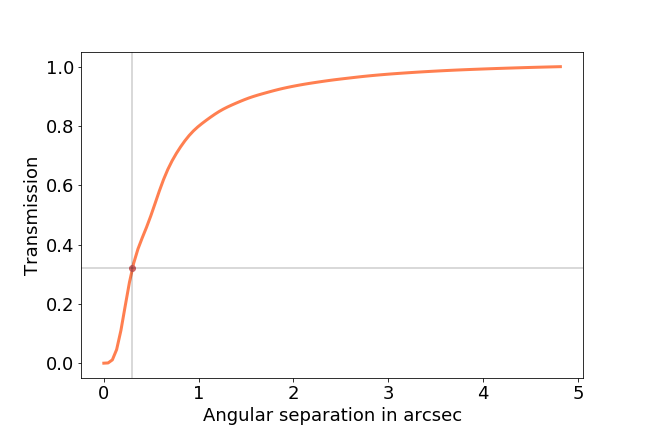}
\caption{Transmission of the AGPM coronagraph as a function of the separation, simulated by scanning a point source radially with respect to the center of the coronagraph. The transmission is measured in the central pixel of the point source image and normalized to the unattenuated PSF. 
The transmission at the expected position of the planet $\beta$ Pic b, at 0.3", is 32\%, identified as a brown dot in the plot. }
\label{fig:transmission}
\end{figure}

\end{appendix}

\end{document}